\documentclass{ws-procs961x669}            
\usepackage{amsfonts}
\usepackage{amsmath}
\usepackage{amssymb}
\usepackage{graphicx}

\newcommand{\be}{\begin{equation}}
\newcommand{\ee}{\end{equation}}
\newcommand{\bal}{\begin{align}}
\newcommand{\eal}{\end{align}}

\begin{document}
\title{
   Lattice QCD : Flavor Physics and Spectroscopy
}

\author{
   Takashi Kaneko$^*$
}

\address{
High Energy Accelerator Research Organization (KEK),
Ibaraki 305-0801, Japan
\\[1mm]
School of High Energy Accelerator Science,
SOKENDAI (The Graduate University for Advanced Studies),
Ibaraki 305-0801, Japan
\\[1mm]
$^*$E-mail: takashi.kaneko@kek.jp
}

\begin{abstract}
  We review highlights of recent results
  on the hadron spectrum and flavor physics from lattice QCD. 
  We also discuss recent rapid progress on the muon anomalous magnetic moment.
\end{abstract}


\bodymatter


\section{Introduction}
\label{sec:intro}




Lattice QCD plays a key role in the intensity frontier
to search for new physics.
Interpretation of experimental measurements
within and beyond the Standard Model (SM)
requires precise knowledge on the relevant hadronic matrix elements,
which describe nonperturbative QCD effects in the underlying processes. 
Moreover, 
high statistics data produced at flavor factories
brought rich outcome about the hadron spectrum, 
such as the discoveries of exotic hadrons. 
Nonperturbative dynamics of QCD characterizes their properties
which do not fit into the simple quark model prediction.
Lattice QCD is the only known method
for ab initio studies of these nonperturbative aspects
with systematically improvable uncertainties.


Lattice QCD is
a regularization of QCD on a discrete Euclidean space-time lattice.
In finite volume, 
the QCD path integral is reduced into a finite-dimensional integral
and can be numerically evaluated by Monte Carlo sampling
of gauge field configurations on a computer.
Pioneering simulations had been limited to
small and coarse lattices with
unphysically heavy and degenerate up and down quarks.
Such limitations have been gradually lifted 
by advances in computing power and
continuous development of lattice QCD formulations and simulation algorithms.
%
%
We note that the lattice formulation is not unique:
the action has a degree of freedom to add irrelevant terms,
which vanish in the limit of zero lattice spacing $a\!\to\!0$,
namely the continuum limit.
We can exploit this freedom to
improve properties of the lattice action
and to firmly establish lattice predictions
by independent calculations with different actions.


Lattice simulations can straightforwardly study
the spectrum and transition amplitude of hadrons
which are stable under QCD.
The light hadron spectrum, for instance,
has been reproduced by simulations
on fine and large lattices 
with light quark masses close to their physical values~\cite{Durr:2008zz,Aoki:2008sm,Bazavov:2009bb,Bietenholz:2011qq}.
While implementation of QED is not straightforward
on a finite periodic space-time volume~\cite{Patella:2017fgk},
even the permille-level neutron--proton mass splitting
is now reproduced by taking account of
electromagnetic (EM) effects and strong isospin breaking
due to the up and down quark mass splitting $m_u\!-\!m_d$~\cite{Borsanyi:2014jba,Horsley:2015eaa}.
The accuracy of simple kaon matrix elements,
namely the decay constant, form factors and bag parameters,
also attains the percent level or better.
Such precision lattice calculation is now being extended 
to heavy-flavored hadrons,
which offer rich probes of new physics.


The study of hadronic decays, however, meets technical difficulties.
As suggested by Maiani\,--\,Testa theorem~\cite{Maiani:1990ca},
there is no simple relation
between the strong decay amplitudes and
correlation functions on the Euclidean lattice.
Theoretical frameworks are under active development
in order to study the $K\!\to\!\pi\pi$ decay
as well as the heavy quarkonia and exotic states,
which generally lie above thresholds.


In this article, we review highlights of recent lattice studies
on the hadron spectrum and flavor physics.
We also briefly discuss recent rapid progress
on the muon anomalous magnetic moment,
which is a key quantity in the search of new physics.
More detailed reviews on these topics can be found
in Refs.\!\citenum{Liu:2016kbb,Wilson:2016rid,Feng:2017voh,Lehner:2017kuc,Colangelo:2017urn}.


\section{Hadron spectroscopy}
\label{sec:spectrum}


If a hadron $H$ is stable under QCD,
it is straightforward to calculated its energy $E_H$ on the lattice.
We prepare
an interpolating field ${\mathcal O}_{H}$ with quantum numbers of $H$,
and extract $E_H$ from the asymptotic behavior of the two-point function
towards the large Euclidean temporal separation $t\!\to\!\infty$
\be
   \langle {\mathcal O}_H(t) {\mathcal O}_H^\dagger(0) \rangle
   \to
   \frac{|Z_H|^2}{2E_H}e^{-E_H t}.
   \label{eqn:spec:2pt:stable}
\ee
Here $Z_H\!=\!\langle 0 | {\mathcal O}_H | H \rangle$
represents the overlap of ${\mathcal O}_{H}$
with the physical state $|H\rangle$.
This simple method forms a basis for the recent postdictions
for the low-lying hadron spectrum mentioned in Sec.~\ref{sec:intro}.


\begin{figure}[t]%
\begin{center}
  \includegraphics[width=0.48\linewidth]{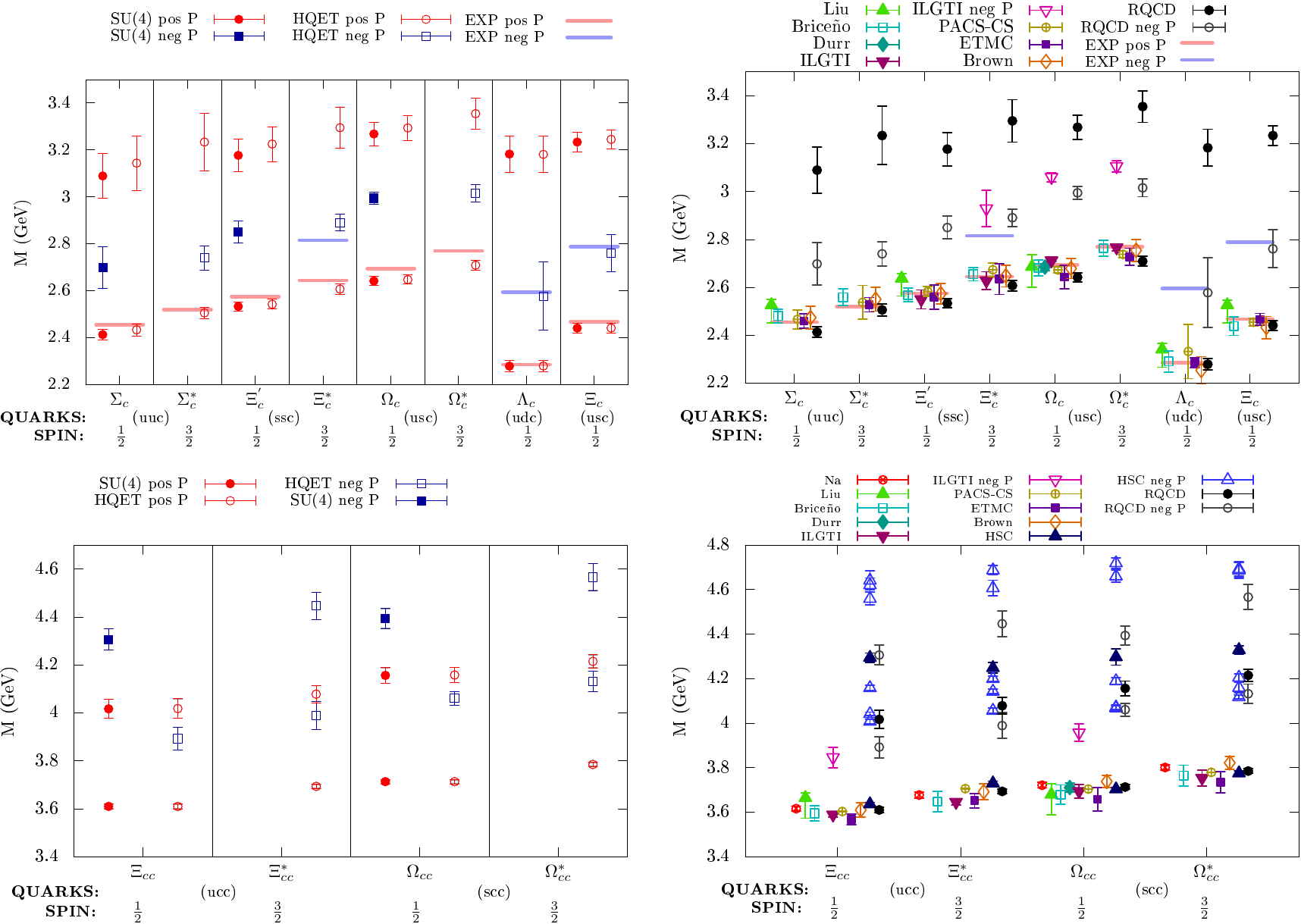}
  \hspace*{2mm}
  \includegraphics[width=0.48\linewidth]{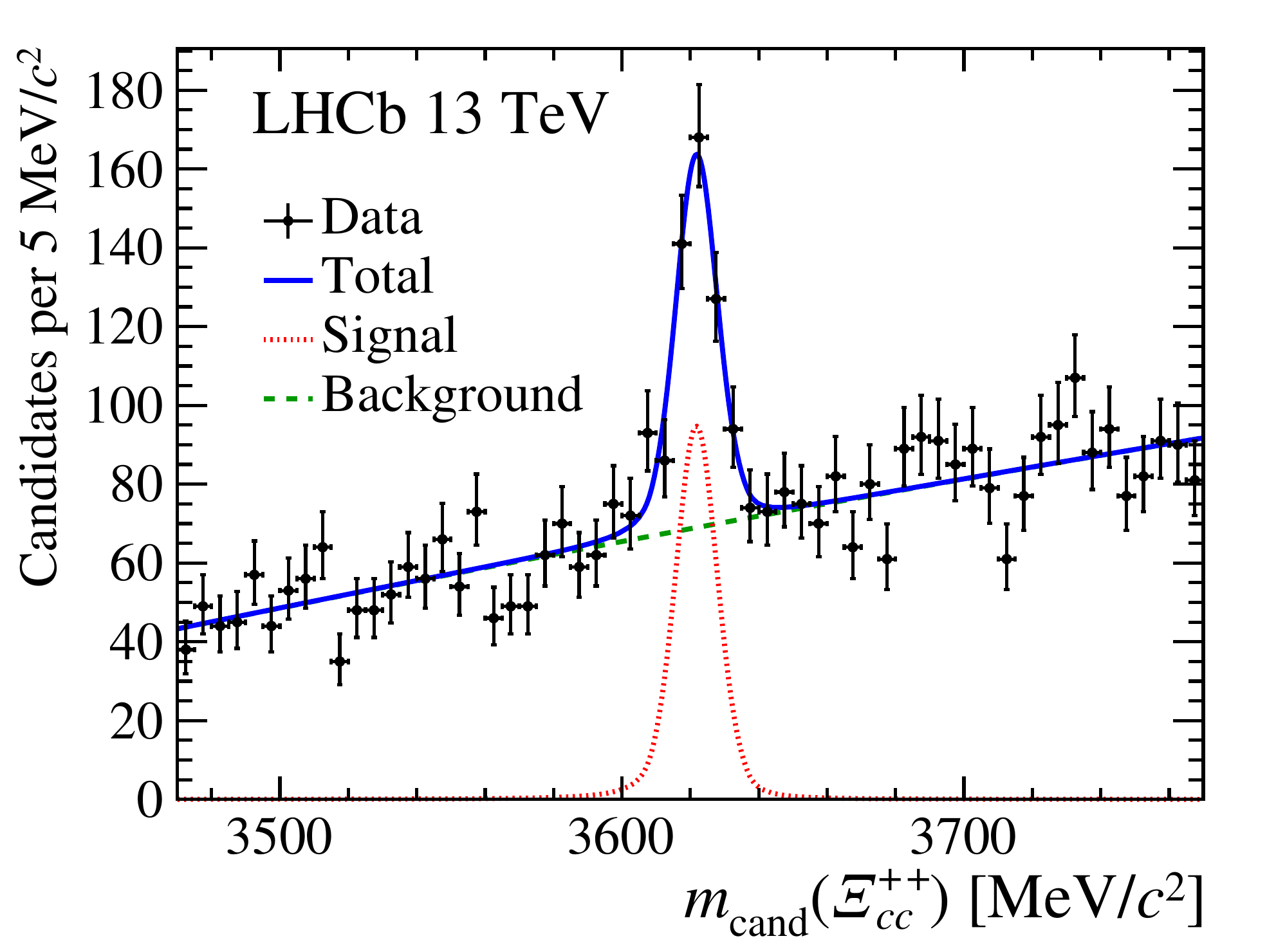}
  \caption{
    Left panel:
    compilation of recent lattice predictions for
    doubly charmed baryon spectra (figure from Ref.\!\citenum{Bali:2015lka}).
    The lowest positive-parity states have been studied by many groups~\cite{Na:2007pv,Liu:2009jc,Basak:2012py,Durr:2012dw,Namekawa:2013vu,Brown:2014ena,Padmanath:2015jea,Bali:2015lka},
    whereas less results are available for the excited and/or negative-parity states~\cite{Basak:2012py,Padmanath:2015jea,Bali:2015lka}.
    Right panel:
    invariant mass distribution of
    $\Xi_{cc}^{++}\!\to\!\Lambda_c^+K^-\pi^+\pi^+$ decay candidates
    from LHCb
    (figure from Ref.\!\citenum{Aaij:2017ueg}).
    $pp$ data sample corrected at a center-of-mass (CM) energy 13 TeV
    with an integrated luminosity of 1.7~fb$^{-1}$ is analyzed.
    The dotted, dashed and solid lines are fit curves for
    the signal, background and their total.
  }
  \label{fig:spec:xicc}
\end{center}
\end{figure}

This can also provide illuminating insight into the nature
of yet-unestablished states which are stable under the strong interaction.
The doubly charmed baryons $\Xi_{cc}^+$ and $\Xi_{cc}^{++}$, for instance,
are expected to have large branching fraction
to flavor changing decay modes~\cite{Kiselev:1998sy}.
The first observation was reported by the SELEX experiment
at $M_{\Xi_{cc}^+}\!=3519(2)$~MeV~\cite{Mattson:2002vu,Ocherashvili:2004hi}
and $M_{\Xi_{cc}^{++}}\!\sim\!3460$~MeV~\cite{Russ:2002bw}.
This however poses a puzzle of the large isospin splitting
$M_{\Xi_{cc}^{++}}-M_{\Xi_{cc}^+}\!\sim\!-60$~MeV and short life-times
in contrast to
phenomenological analyses~\cite{Hwang:2008dj,Guberina:1999mx,Kiselev:1998sy}.
The left-panel of Fig.~\ref{fig:spec:xicc} is a compilation of
recent lattice QCD predictions for the doubly-charmed baryon spectra
in the isospin limit~\cite{Na:2007pv,Liu:2009jc,Basak:2012py,Durr:2012dw,Namekawa:2013vu,Brown:2014ena,Padmanath:2015jea,Bali:2015lka}.
These studies employing different lattice actions 
have led to reasonable agreement around $M_{\Xi_{cc}}\!\sim\!3600$~MeV,
which is systematically higher than the SELEX results
but favors recent observation $M_{\Xi_{cc}^{++}}\!=\!3621.40(0.78)$~MeV
by LHCb (right panel of Fig.~\ref{fig:spec:xicc}).
A lattice estimate of the isospin splitting
$M_{\Xi_{cc}^{++}}\!-\!M_{\Xi_{cc}^+}\!=\!2.2(0.2)$~MeV
also contradicts the old measurement.


While studying unstable particles is more involved,
there has been considerable progress in recent years.
Let us consider a resonance 
strongly decaying into two particles $A$ and $B$.
Its mass and width can be determined from
the scattering amplitude of $A$ and $B$,
but they are not directly given by the amplitudes of
the Euclidean correlation function~\cite{Maiani:1990ca}
\be
   \langle {\mathcal O}_{AB}^{\prime}(t) {\mathcal O}_{AB}^\dagger(0) \rangle
   = 
   \sum_{n} 
   \frac{Z_{AB,n}^\prime Z_{AB,n}^*}{2E_{AB,n}}e^{-E_{AB,n} t}.
   \label{eqn:spec:2pt:scattering}
\ee
Here ${\mathcal O}_{AB}$ and ${\mathcal O}_{AB}^{\prime}$ represent
interpolating fields for the two-particle state $AB$,
and $n$ is the index of the energy levels.
However,
the spectrum $\{E_{AB,1},E_{AB,2},...\}$
deviates from the non-interacting energy levels
due to the $AB$ scattering on the lattice, 
and hence encodes the scattering amplitude.
L\"uscher derived a formula to extract the phase shift
from the discrete spectrum on the finite volume~\cite{Luscher:1986pf,Luscher:1990ux,Luscher:1991cf}.
%
Later the HALQCD Collaboration developed another method to
study a multi-particle system~\cite{Ishii:2006ec,HALQCD:2012aa,Aoki:2012bb}.
By using an interpolating field
$O_{AB}^\prime(t)\!=\!O_A({\bf x},t)O_B({\bf x}+{\bf r},t)$,
the HALQCD method extracts the Nambu-Bethe-Salpeter wave function
of the two-particle system $AB$ with relative distance $|{\bf r}|$, 
from which the amplitude of the $AB$ scattering can be deduced.


\begin{figure}[t]%
\begin{center}
  \includegraphics[width=0.412\linewidth]{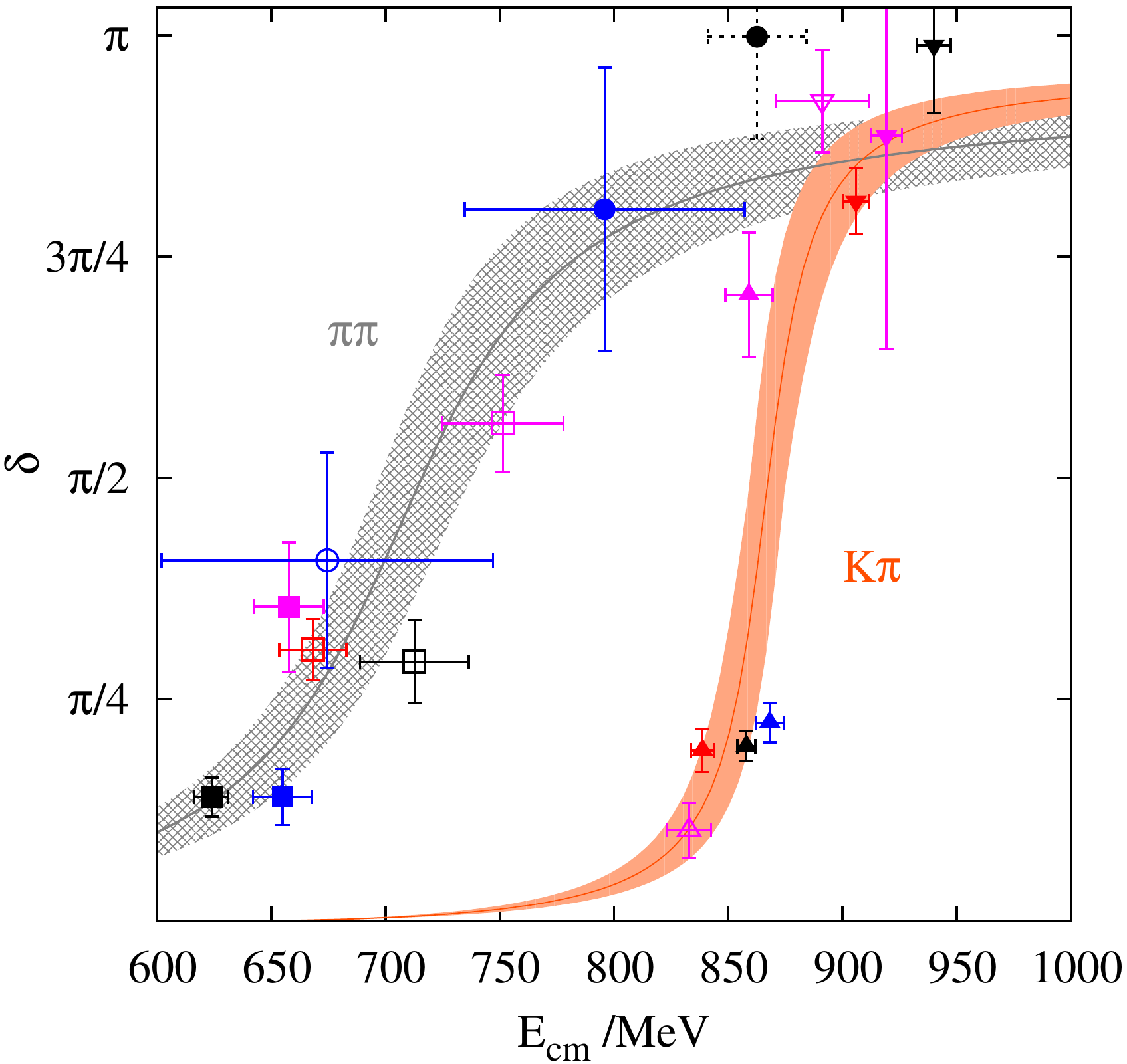}
  \hspace*{1mm}
  \includegraphics[width=0.562\linewidth]{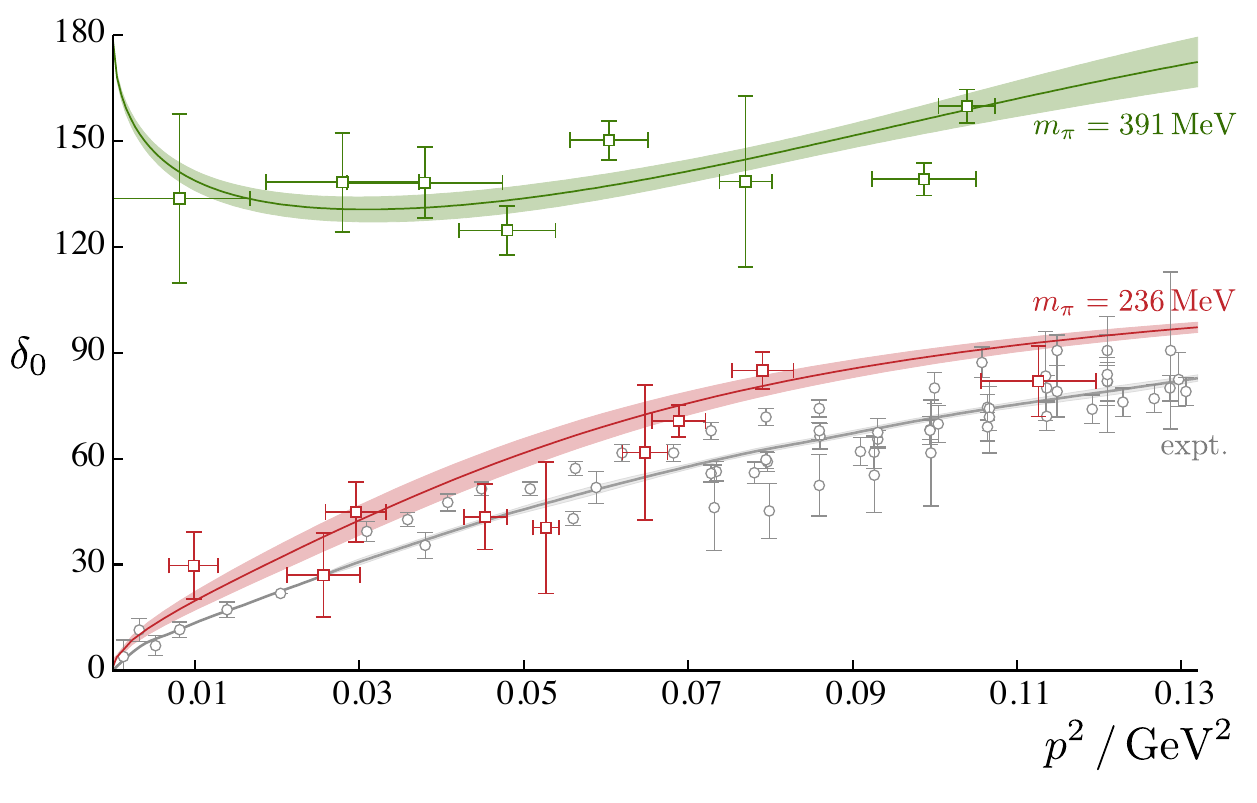}
  \caption{
    Left panel:
    isovector $\pi\pi$ ($\delta_{1}^{\pi\pi}$, circles and squares) and
    $K\pi$ ($\delta_{1}^{K\pi}$, triangles) phase shift
    against the CM frame energy
    (figure from Ref.\!\citenum{Bali:2015gji}).
    The black shaded and orange bands show
    the Breit-Wigner parametrization of $\delta_{1}^{\pi\pi}$ and
    $\delta_{1}^{K\pi}$, respectively.
    Right panel:
    isoscalar $\pi\pi$ scattering phase shift $\delta_{0}^{\pi\pi}$
    as a function of scattering momentum $p^2\!=\!(E_{\rm CM}/2)^2-M_\pi^2$
    (figure from Ref.\!\citenum{Briceno:2016mjc}).
  }
  \label{fig:spec:pipi}
\end{center}
\end{figure}

A good application of these methods is rigorous understanding of
the resonant $\pi\pi$ and $K\pi$ scatterings directly from QCD.
While the original L\"uscher formula was limited to a single channel problem
for two identical scalar particles in the CM frame,
efforts over the past few decades have generalized it
to arbitrary two-particle systems
also in moving frames~\cite{Rummukainen:1995vs,Kim:2005gf,Christ:2005gi,Lage:2009zv,He:2005ey,Briceno:2014oea}.
This theoretical development has made rapid stride in  
the lattice study of the $\rho$ and $K^*$ resonances in the isovector channel~\cite{Liu:2016kbb,Wilson:2016rid,Thomas:2017fnb}. 
The left panel of Fig.~\ref{fig:spec:pipi} shows the scattering phase shift
$\delta_{1}^{\pi\pi(K\pi)}$ obtained by the RQCD Collaboration.
They simulate a pion mass $M_\pi\!\sim\!150$~MeV close to
its physical value $M_{\pi,\rm phys}$ in two-flavor QCD~\cite{Bali:2015gji},
where only degenerate up and down quarks are present in the sea.
It is encouraging to observe the rapid raise of $\delta_{1}^{\pi\pi(K\pi)}$
near the physical $\rho$ ($K^*$) mass.
Although a detailed analysis based on a Breit-Wigner parametrization
leads to slight tension in the resonance masses and widths with experiment,
it may be attributed to quenching of strange quarks,
namely the missing $K\bar{K}$ channel~\cite{Guo:2016zos,Hu:2016shf}.
Calculations in three-flavor QCD,
namely with  dynamical strange quarks,
are being available.
For recent studies,
see Refs.~\citenum{Wilson:2015dqa,Fu:2016itp,Alexandrou:2017mpi}
and references therein.


The L\"uscher method has been also applied to the isoscalar channel,
which is much more challenging than the iso-nonsinglet ones.
The relevant two-point function involves the so-called
quark-disconnected diagrams (see Fig.~1 of Ref.\!\citenum{Briceno:2016mjc}),
which are computationally very expensive.
The Hadron Spectrum Collaboration have published the first full
calculation of the isoscalar scattering phase shift
$\delta_{0}^{\pi\pi}$~\cite{Briceno:2016mjc}.
It is interesting to observe that 
$\sigma$ is a bound state at $M_\pi\!\sim\!390$~MeV,
but turns into a broad resonance already at $\!\sim\!240$~MeV
as seen in the right panel of Fig.~\ref{fig:spec:pipi}.
The mass and width of $\sigma$ approach
their experimental values~\cite{Pelaez:2015qba}, as $M_\pi$ decreases.
While a simulation at the physical mass $M_{\pi,\rm phys}$
is needed to make a direct comparison with experiment,
those at unphysical $M_\pi$'s deepen our understanding of
the existence form of the hadrons:
how it changes from unphysical to the real worlds.


Extension to heavy hadrons,
particularly exotic hadrons discovered at flavor factories, 
is intriguing but still challenging task of lattice QCD.
These hadrons in general have significant branching fraction
to states containing three or more particles.
Generalization of L\"uchser's framework capable of
these high multiplicity states is an active area of lattice QCD~\cite{Polejaeva:2012ut,Hansen:2012tf,Briceno:2012yi,Briceno:2017tce}.
It is not unreasonable to hope that such a general framework
will become available in the next decade.

\begin{figure}[t]%
\begin{center}
  \includegraphics[width=0.485\linewidth,clip]{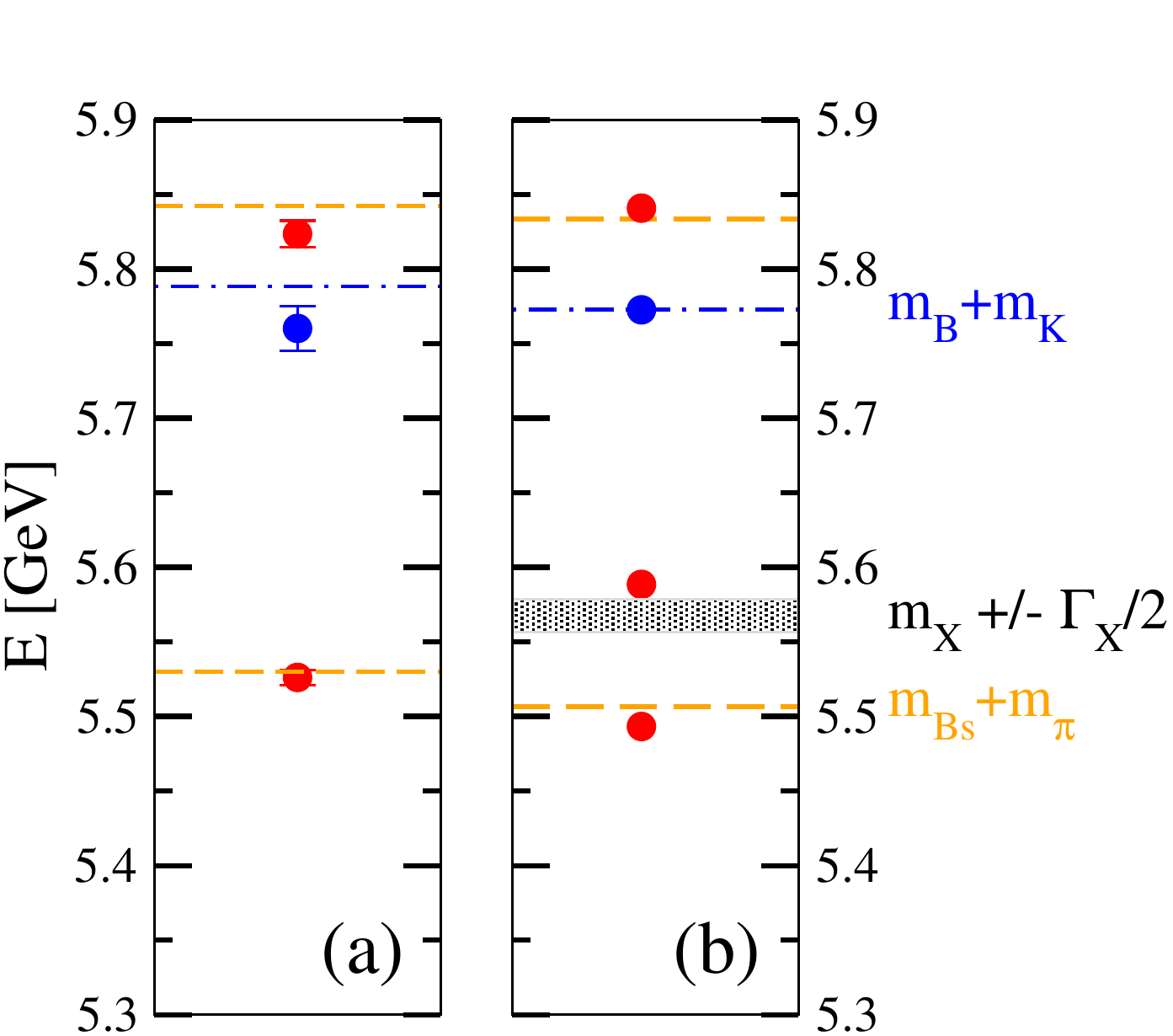}
  \hspace*{1mm}
  \includegraphics[width=0.485\linewidth,clip]{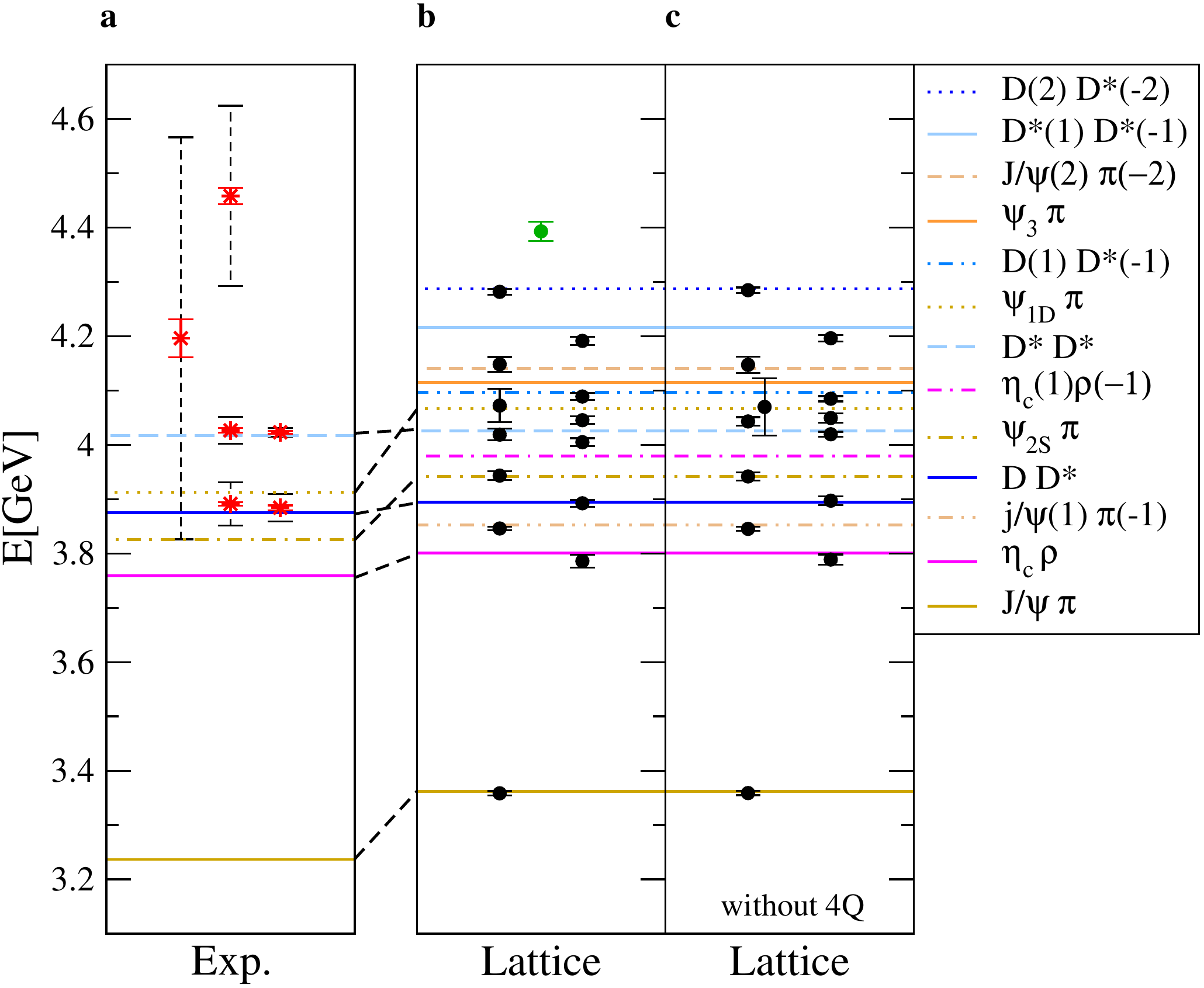}
  \caption{
    Left panel:
    finite volume spectrum for $X(5568)$
    (figure from Ref.\!\citenum{Lang:2016jpk}).
    The sub-panel (a) shows simulation data at $M_\pi\!\sim\!156$~MeV,
    whereas an analytic prediction at $M_{\pi,\rm phys}$ is plotted
    in the sub-panel (b) by assuming the existence of $X(5568)$.
    The finite volume spectrum ($L\!=\!2.9$~fm) is plotted by circles.
    The horizontal lines show non-interacting levels of
    $B_s(0)\pi^+(0)$, $B^+(0)\bar{K}^0(0)$ and $B_s(1)\pi^+(-1)$,
    where the arguments are lattice momenta in units of $2\pi/L$.
    Note that
    the finite volume level (red circle) near the experimental mass
    (shaded band in sub-panel (b))
    is missing in the simulation data (sub-panel (a)).
    Right panel:
    finite volume spectrum for $Z_c^+(3900)$
    (figure from Ref.\!\citenum{Prelovsek:2014swa}).
    The experimental masses of the $Z_c$ candidates discussed in Ref.\!\citenum{Brambilla:2014jmp}
    ($Z_c(3885,3900,4020,4025,4430)$) are plotted in the sub-panel ``a''.
    The vertical dashed lines represent twice the widths.
    The sub-panels ``b'' and ``c'' show finite volume spectrum
    at $M_\pi\!=\!266$~MeV obtained with different choices
    of the lattice interpolating fields.
    We note that black symbols are identified as scattering states
    shown by horizontal lines,
    and there is not extra energy level corresponding to $Z_c^+(3900)$.
  }%
  \label{fig:spec:exotic:luescher}
\end{center}
\end{figure}


There are however good examples that the current methodology
can gain insight into the nature of exotic hadrons.
One example is a recent study of $X(5568)$~\cite{Lang:2016jpk}.
The D0 Collaboration recently reported
a narrow peak ($\Gamma\!\sim\!22$~MeV) in the $B_s\pi^+$ invariant mass
of their $p\bar{p}$ collision data~\cite{D0:2016mwd},
while the later LHCb measurements did not confirm
the peak structure~\cite{Aaij:2016iev}.
This state, if exists, has an interesting exotic content
with four different quark flavors $\bar{b}s\bar{d}u$.
It strongly decays into a two-particle state $B_s\pi^+$
and lies significantly below other thresholds. 
Therefore the L\"uscher formula can be applied
rather straightforwardly.
Lang {\it et al.} calculated the finite volume spectrum
in Eq.~(\ref{eqn:spec:2pt:scattering})
by numerical simulation near $M_{\pi,\rm phys}$.
The left panel of Fig.~\ref{fig:spec:exotic:luescher} compares it 
with an analytic estimate at $M_{\pi,\rm phys}$ 
assuming the existence of $X(5568)$.
The energy level corresponding to $X(5568)$ is missing 
in the simulation data.
This disfavors the existence of $X(5568)$
in agreement with the LHCb measurement.

\begin{figure}[t]%
\begin{center}
  \includegraphics[width=0.494\linewidth]{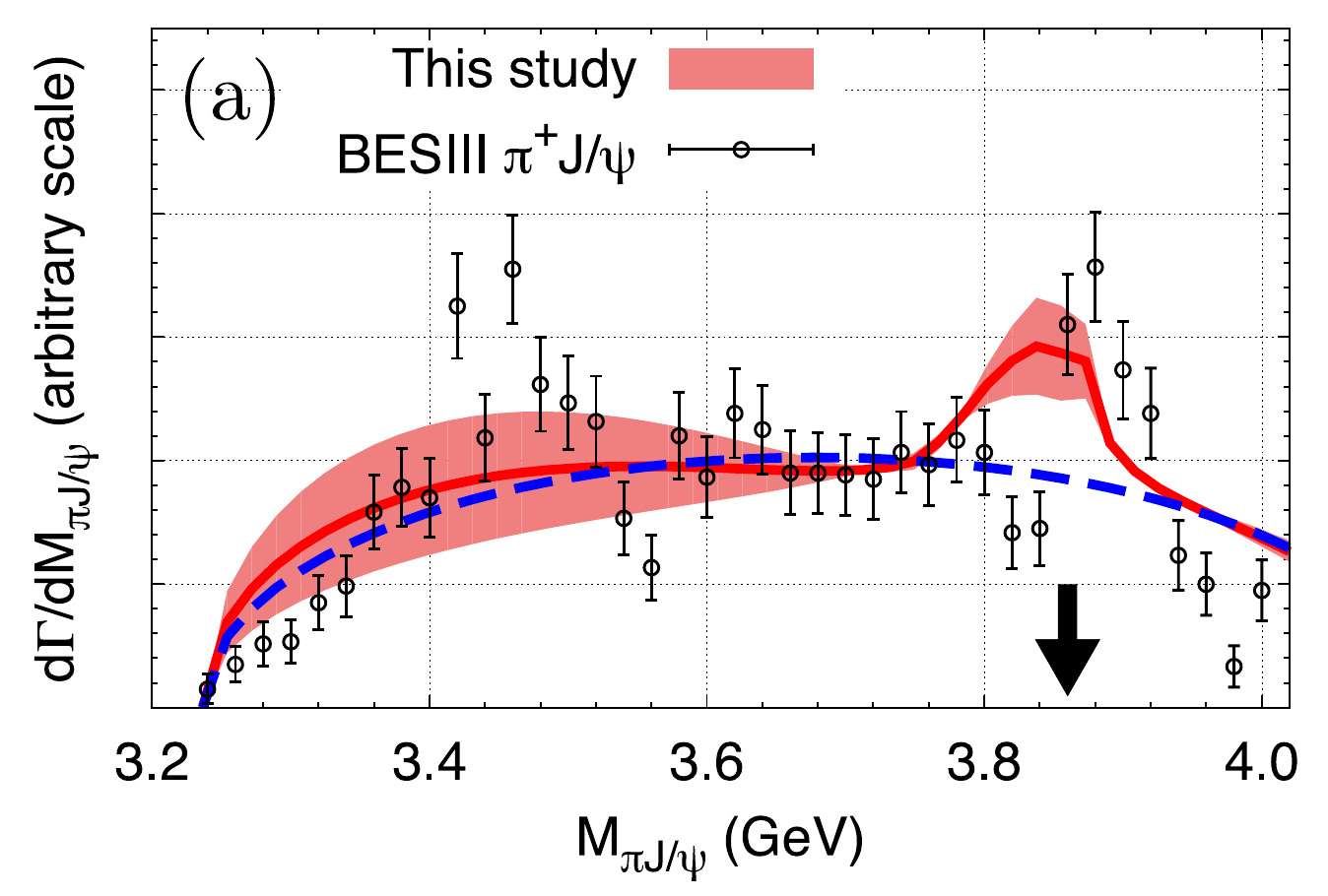}
  \hspace*{1mm}
  \includegraphics[width=0.478\linewidth]{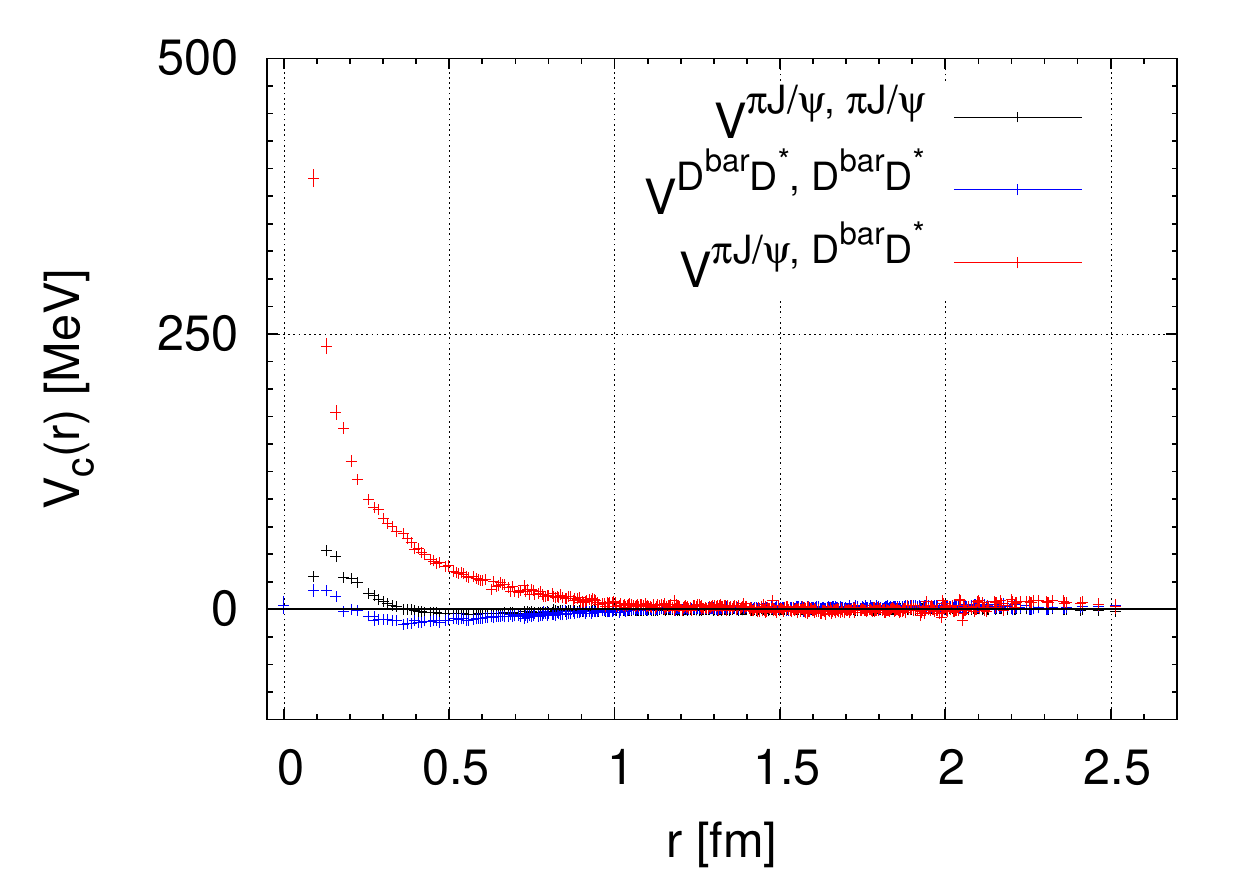}
  \caption{
     Right panel:
     $J/\Psi\pi^+$ invariant mass distribution
     (figure from Ref.~\citenum{Ikeda:2016zwx}).
     Open circles are BESIII data~\cite{Ablikim:2013mio}.
     The vertical arrow indicates the $Z_c$ peak position.
     The red band shows 
     a theoretical estimate obtained from the scattering matrix
     by the HALQCD method.
     The peak structure disappears in the blue dashed line,
     which is obtained by turning off the off-diagonal potential.
     Left panel:
     coupled-channel potential among $J/\psi\pi$ and $\bar{D}D^*$ states
     (courtesy of Yoichi Ikeda (HALQCD Collaboration)).
     The black and blue symbols show the diagonal potentials of
     the $J/\psi\pi$ and $\bar{D}D^*$ states, respectively.
     The off-diagonal potential between these states is plotted
     by the red symbol.
  }%
  \label{fig:spec:exotic:halqcd}
\end{center}
\end{figure}


There has also been recent interesting progress for $Z_c^+(3900)$,
another candidate for four quark exotic states.
BESIII reported a peak structure slightly above the $\bar{D}D^*$ threshold 
in the $J/\psi \pi^+$ invariant mass distribution~\cite{Ablikim:2013mio},
which was also confirmed in Belle and CLEO-c data~\cite{Liu:2013dau,Xiao:2013iha}.
The finite volume spectrum has already been studied~\cite{Prelovsek:2014swa,Chen:2014afa,Lee:2014uta}. 
As shown in the right panel of Fig.~\ref{fig:spec:exotic:luescher},
any extra energy level corresponding to $Z_c$ has not been confirmed
suggesting the possibility of the $Z_c$ peak of kinematical origin.
Recently, the scattering matrix among three states,
$J/\psi$, $\bar{D}D^*$ and $\rho\eta_c$, has been determined
at unphysical $M_\pi\!\gtrsim\!410$~MeV by the HALQCD method~\cite{Ikeda:2016zwx}.
As shown in the left panel of Fig.~\ref{fig:spec:exotic:halqcd},
it reproduces the $Z_c$ peak structure,
which however disappears by turning off the strong coupling between $J/\psi\pi$
and $\bar{D}D^*$ shown in the right panel.
This suggests that the $Z_c$ peak is a threshold cusp effect
due to the opening of the $\bar{D}D^*$ threshold.
This interesting observation should be confirmed at $M_{\pi,\rm phys}$,
since unphysically heavy $M_\pi$ significantly disorders
possibly relevant thresholds $\psi_{\{2S,1D,3\}}\pi$.
A coupled-channel analysis using the L\"uscher formula
is also welcome to firmly establish the origin of the $Z_c$ peak structure.


\section{Flavor physics}
\label{sec:flavor}


There has been steady progress in precision lattice study of 
the (semi)leptonic decays and neutral meson mixings. These processes
provide determination of relevant Cabibbo-Kobayashi-Maskawa (CKM)
matrix elements 
and a wealth of probes of new physics,
such as the fully differential decay distribution. 
For processes where we can safely ignore the final state interaction,
the relevant hadronic matrix elements and hadronic inputs
can be straightforwardly calculated from correlation functions on the lattice.
For the kaon leptonic decay, for instance,
the amplitude $Z_H$ of the two-point function (\ref{eqn:spec:2pt:stable})
of the weak axial current ${\mathcal O}_H\!=\!A_4$
gives the matrix element $\langle 0 | A_4 | K(p) \rangle\!=\!M_K f_K$,
which is parametrized by the kaon decay constant $f_K$.
Form factors for semileptonic decays
and bag parameters for neutral meson mixings can be similarly extracted
from tree-point functions.
This is rather straightforward procedure, and hence 
good simulation setup, namely the choice of the lattice formulation
and simulation parameters, is a key to perform a high precision calculation.

\begin{figure}[t]%
\begin{center}
  \includegraphics[width=0.388\linewidth]{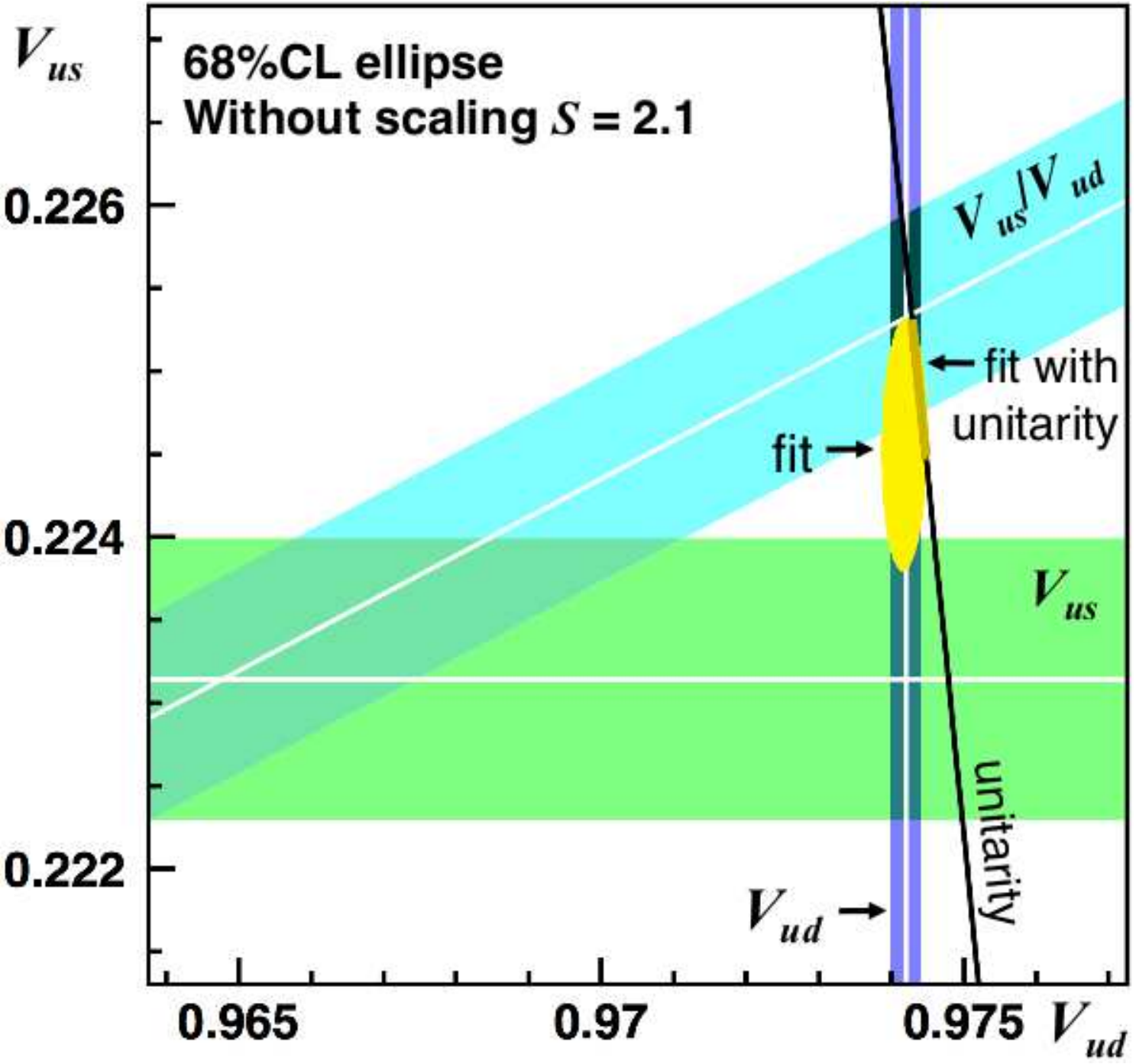}
  \hspace*{1mm}
  \includegraphics[width=0.587\linewidth]{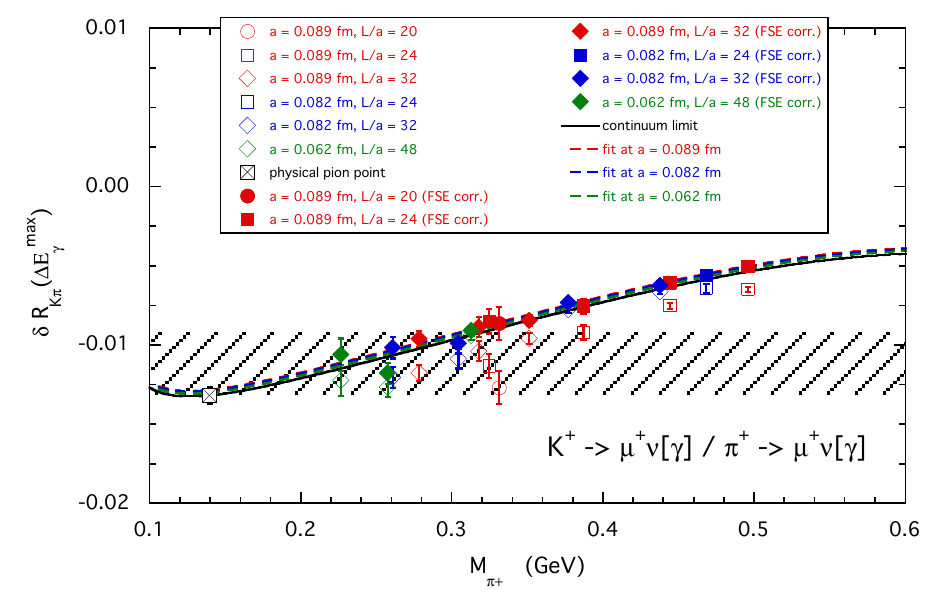}
  \caption{
     Left panel:
     test of CKM unitarity in $(|V_{ud}|,|V_{us}|)$ plane
     (figure from Ref.\!\citenum{Moulson:2017ive}).
     The horizontal and oblique bands indicate
     $|V_{us}|$ determined from the $K\!\to\!\pi\ell\nu$ decay
     and $|V_{us}/V_{ud}|$ from the kaon and pion leptonic decays, respectively.
     The vertical band is $|V_{ud}|$
     from the super-allowed nuclear $\beta$ decays.
     A fit to these data yields $(|V_{ud}|,|V_{us}|)$ shown in the yellow region.
     The black solid line satisfies CKM unitarity in the first row,
     where $|V_{ub}|$ has small effects.
     Right panel: isospin correction $\delta R_{K\pi}$ 
     to $\Gamma(K\to\ell\nu)/\Gamma(\pi\to\ell\nu)$
     (figure from Ref.\!\citenum{Lubicz:2016mpj}).
     Symbols show lattice data,
     whereas the dashed and solid lines are their fit curves
     at finite lattice spacings and in the continuum limit,
     respectively.
     The shaded band represents
     the ChPT estimate~\cite{Cirigliano:2011tm,Rosner:2015wva}.
  }%
  \label{fig:flv:kaon:semi+leptonic}
\end{center}
\end{figure}

There have been many independent studies on sufficiently large and fine
lattices near $M_{\pi, \rm phys}$ for the kaon (semi)leptonic decay and mixing.
For instance, the current accuracy reaches the sub\,\% level
for the $K\!\to\!\pi$ form factor and
the ratio of the kaon and pion decay constants $f_K/f_\pi$~\cite{Aoki:2016frl}.
These hadronic inputs are used to precisely determine 
$|V_{us}|$ and $|V_{us}/V_{ud}|$
as shown in the left panel of Fig.~\ref{fig:flv:kaon:semi+leptonic}.
Together with $|V_{ud}|$ from the super-allowed nuclear $\beta$ decays,
CKM unitarity in the first row 
$|V_{ud}|^2+|V_{us}|^2+|V_{ub}|^2\!=\!1$
is now confirmed at the 0.1\,\% level~\cite{Kaneko:2017ysl}.
This is one of the most precise unitarity test,
and can probe the new physics scale up to $\approx\!10$~TeV~\cite{Cirigliano:2009wk,Gonzalez-Alonso:2016etj}.

At the impressive accuracy of the hadronic inputs,
the uncertainty of the EM and strong isospin breaking corrections
to the decay rate is  no longer negligible. 
These corrections have been conventionally estimated
in chiral perturbation theory (ChPT)
with typical accuracy of 0.2\,--\,0.4\,\%~\cite{Antonelli:2010yf}.
It is however difficult to improve the ChPT calculation
by extending to higher orders,  
where many additional unknown low energy constants appear.
Lattice QCD calculation of these corrections
takes on increasing importance 
and is being actively pursued~\cite{Patella:2017fgk}.
The presence of the infrared divergences
complicates the calculation of the EM corrections
for the (semi) leptonic decays.
Recently,
a new strategy was proposed~\cite{Carrasco:2015xwa}
and successfully applied to the leptonic decay rate ratio
$\Gamma(K\!\to\!\ell\nu)/\Gamma(\pi\!\to\!\ell\nu)$,
which provides the determination of $|V_{us}|/|V_{ud}|$.
As shown in the right panel of Fig.~\ref{fig:flv:kaon:semi+leptonic},
their preliminary estimate
$\delta R_{K\pi}\!=\!-0.0137(13)$~\cite{Lubicz:2016mpj}
is in good agreement with the conventional ChPT estimate -0.0112(21)~\cite{Cirigliano:2011tm,Rosner:2015wva}.
The lattice estimate is, however, systematically improvable
by more realistic simulations, for instance, 
near $M_{\pi,\rm phys}$ with higher statistics.



Another thrust of recent lattice efforts in kaon physics is
application to more involved processes,
such as the $K\!\to\!\pi\pi$ hadronic decay and rare decays.
Similar to the scattering amplitudes discussed in Sec.~\ref{sec:spectrum},
the $K\!\to\!\pi\pi$ decay amplitudes are not directly given by 
the correlation functions on the lattice.
Lellouch and L\"uscher derived a formula to relate the finite volume
matrix elements to the physical amplitudes~\cite{Lellouch:2000pv}.
This has been successfully implemented in recent works
by the RBC/UKQCD Collaborations
for the $\Delta I\!=\!3/2$~\cite{Blum:2015ywa}
and 1/2~\cite{Bai:2015nea} channels.
They obtain the direct CP violation parameter
${\rm re}\left[ \epsilon^\prime/\epsilon \right]
\!=\!1.4(5.2)_{\rm stat}(4.6)_{\rm sys}\!\times\!10^{-4}$.
A less precise but consistent value $8(25)_{\rm stat}\!\times\!10^{-4}$
was obtained by an independent calculation
with a different lattice formulation~\cite{Ishizuka:2015oja}.
A slight tension with the experimental value~\cite{AlaviHarati:2002ye,Batley:2002gn,Abouzaid:2010ny}
$16.6(2.3)\!\times\!10^{-4}$
is of great phenomenological interest~\cite{Buras:2015yba,Kitahara:2016nld}.
The RBC/UKQCD Collaboration is making continuous efforts, which aim
improved statistical accuracy of a factor of two~\cite{Kelly:2018qgs},
and better control of leading systematic uncertainties,
for instance, due to the operator renormalization~\cite{Kelly:2016yqi}.
We can therefore hope an improved estimate of 
${\rm re}\left[ \epsilon^\prime/\epsilon \right]$ in the near future.
Note also that a first exploratory study is available both for
$K\!\to\!\pi\ell^+\ell^-$~\cite{Christ:2016mmq} and 
$K\!\to\!\pi\nu\bar{\nu}$~\cite{Bai:2017fkh} decays.


\begin{figure}[t]%
\begin{center}
  \includegraphics[width=0.487\linewidth,clip]{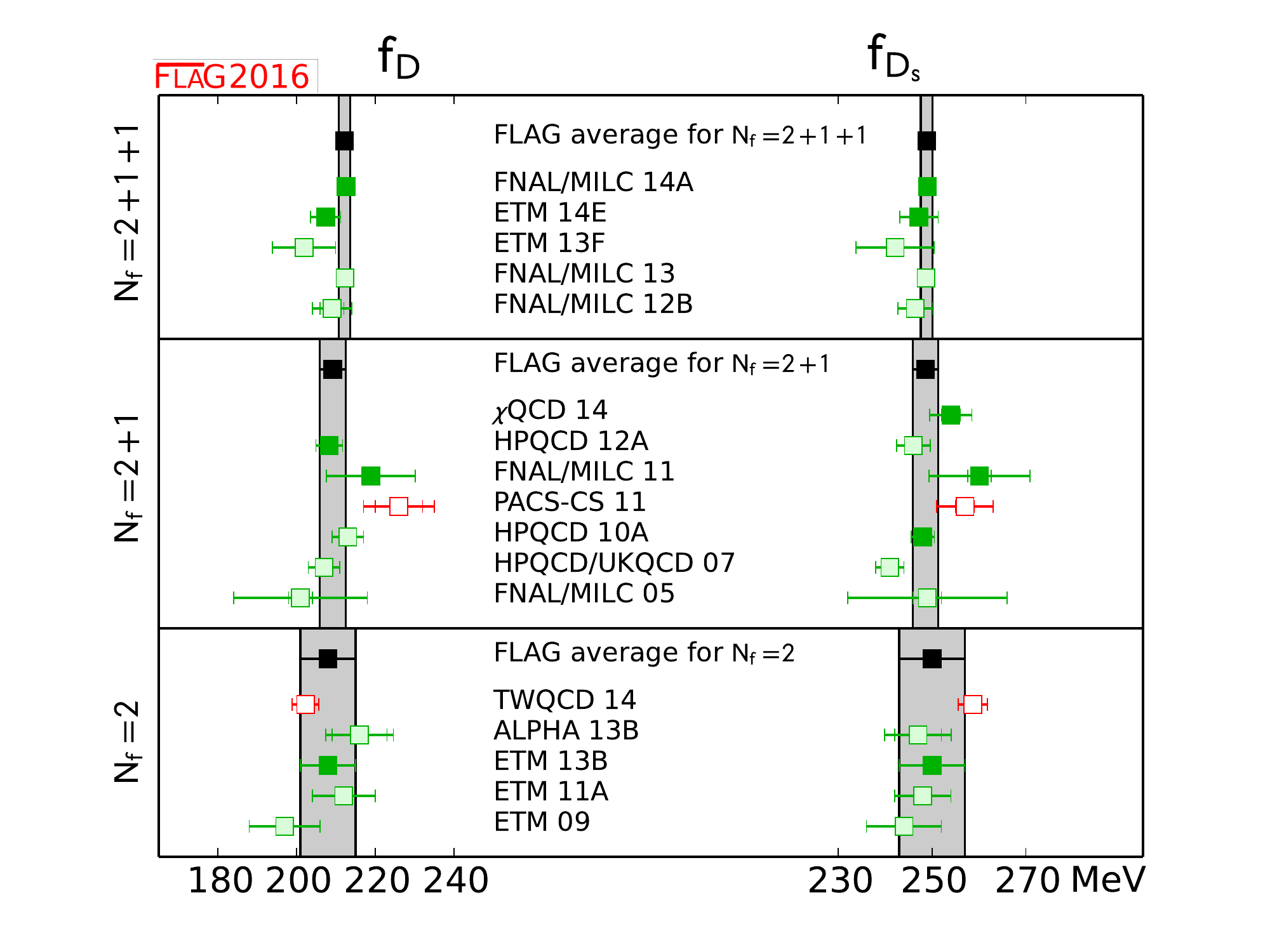}
  \hspace*{1mm}
  \includegraphics[width=0.487\linewidth,clip]{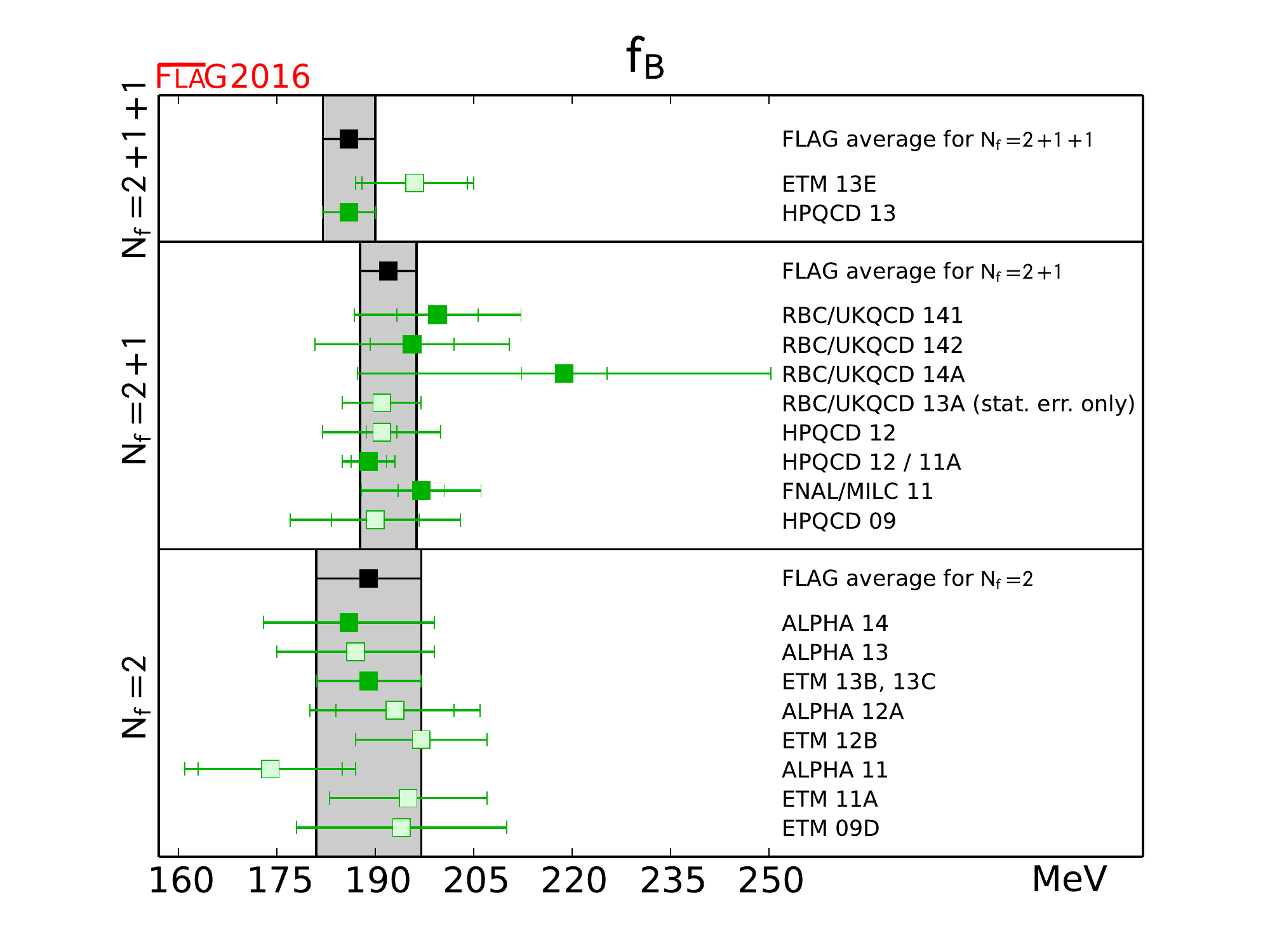}
  \caption{
    Compilation of recent lattice results for
    heavy-light decay constants $f_D$, $f_{D_s}$ (left panel)
    and $f_B$ (right panel)
    (figures from the FLAG review~\cite{Aoki:2016frl}).
    Green symbols are obtained with systematics under control,
    whereas reds are not.
    Black squares are average of the filled symbols at each $N_f$,
    that is the number of flavors of dynamical quarks.
  }%
  \label{fig:flv:heavy:leptonic}
\end{center}
\end{figure}

The accuracy of the heavy-light meson decay constants has been significantly
improved over the past several years.
Figure~\ref{fig:flv:heavy:leptonic} shows that
there have been many independent calculations with systematics under control.
The accuracy of the world average
quoted by the Flavor Lattice Averaging Group (FLAG)~\cite{Aoki:2016frl}
is $\approx$\,0.6\,\% for $f_{D_{(s)}}$ and $\approx$\,2\,\% for $f_{B_{(s)}}$.
These are well below the current experimental precision~\cite{Amhis:2016xyh},
and the isospin correction starts being relevant at this level of accuracy.


\begin{figure}[t]%
\begin{center}
  \includegraphics[width=0.5\linewidth,clip]{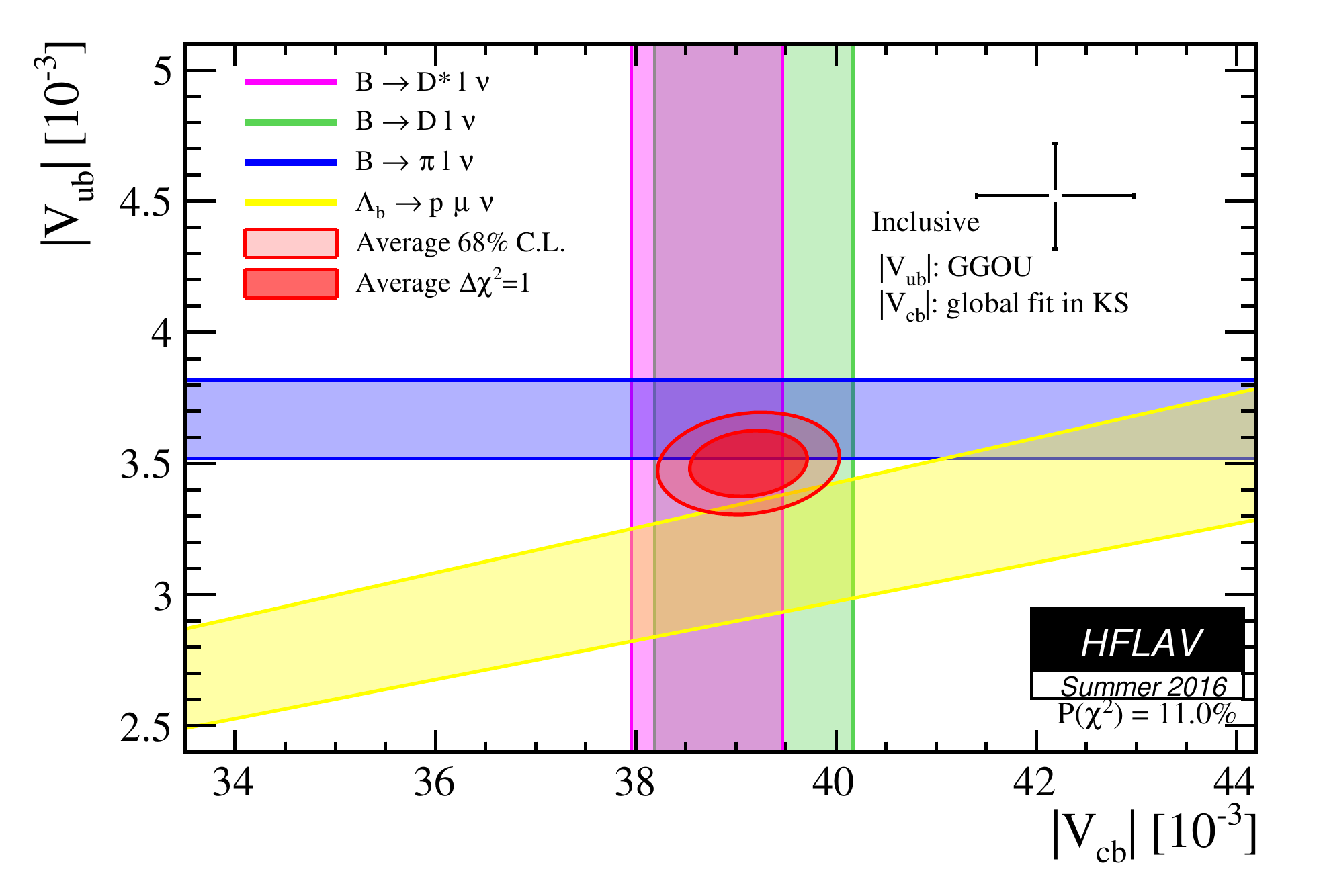}
  \caption{
    $|V_{ub}|$ versus $|V_{cb}|$
    (figure from Ref.\!\citenum{Amhis:2016xyh}).
    Horizontal and vertical
    bands represent $|V_{ub}|$ from $B\!\to\!\pi\ell\nu$
    and $|V_{cb}|$ from $B\!\to\!D^{(*)}\ell\nu$, respectively.
    The oblique band is $|V_{ub}|/|V_{cb}|$ from $\Lambda_b\!\to\!p\ell\nu$ and
    $\Lambda_c\ell\nu$. 
    The average of these estimates is shown by the red region.
    These should be compared with the point with the error bars
    obtained from the inclusive decays $B\!\to\!X_{u(c)}\ell\nu$.
  }%
  \label{fig:flv:heavy:sld:hint}
\end{center}
\end{figure}

%
The $B\!\to\!\pi\ell\nu$ and $B\!\to\!D^{(*)}\ell\nu$ semileptonic decays
provide the conventional determination of the CKM matrix elements
$|V_{ub}|$ and $|V_{cb}|$, respectively.
As shown in Fig.~\ref{fig:flv:heavy:sld:hint},
however, 
there has been a long-standing tension between these exclusive
and inclusive decays~\cite{Amhis:2016xyh}.
While this could be a sign of new physics~\cite{Chen:2008se},
we clearly need more thorough theoretical and experimental studies
to fully resolve/understand the tension.
Lattice QCD plays a crucial role in controlling
the dominant theoretical uncertainty
arising from the relevant hadronic matrix elements.


\begin{figure}[b]%
\begin{center}
  \includegraphics[width=0.495\linewidth,clip]{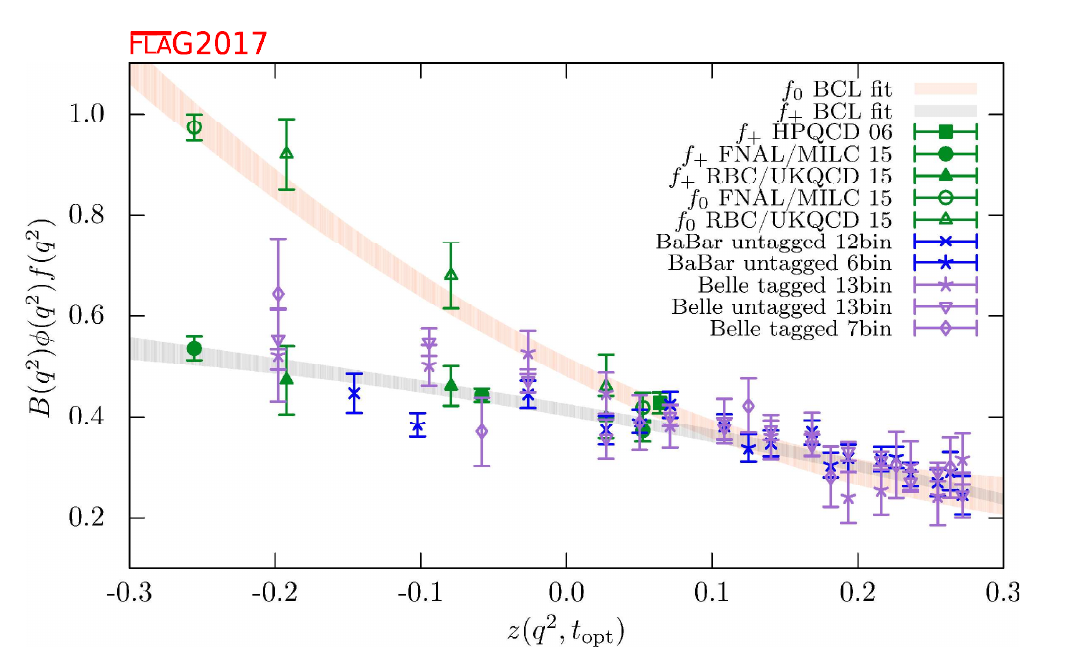}
  \includegraphics[width=0.495\linewidth,clip]{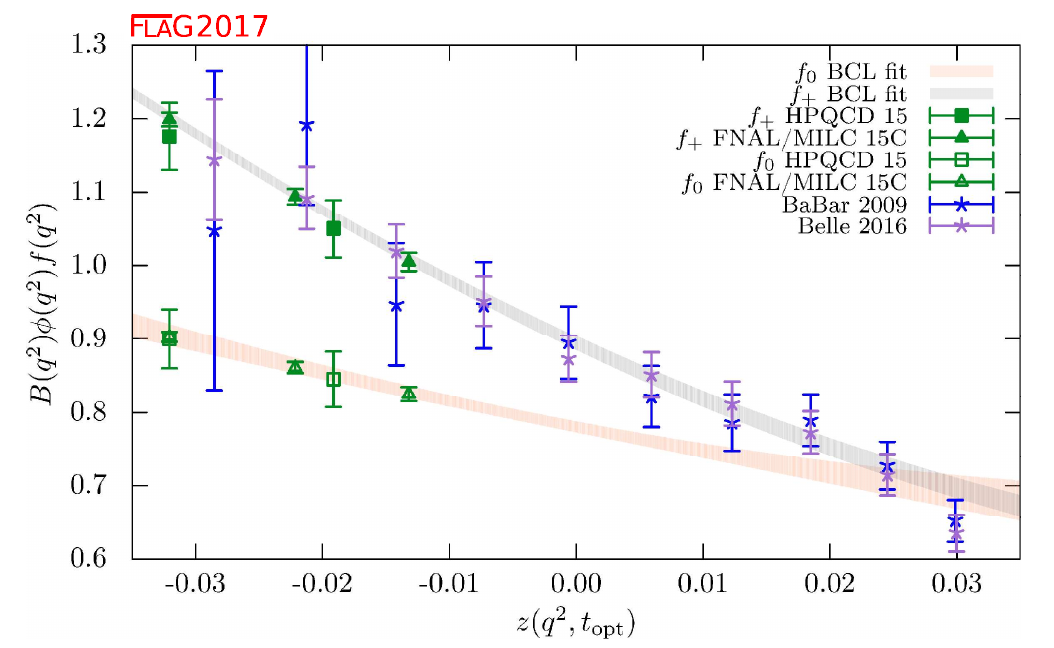}
  \caption{
    Simultaneous fits
    to determine $|V_{ub}|$ from $B\!\to\!\pi\ell\nu$ (left panel)
    and $|V_{cb}|$ from $B\!\to\!D\ell\nu$ (right panel)
    (figures from Ref.\!\citenum{Aoki:2016frl}).
    Recent lattice results for $f_+^{B\pi(BD)}(q^2)$~\cite{Dalgic:2006dt,Flynn:2015mha,Lattice:2015tia,Lattice:2015rga,Na:2015kha}
    and $|V_{ub(cb)}|f_+^{B\pi(BD)}(q^2)$ from
    BaBar~\cite{delAmoSanchez:2010af,Lees:2012vv,Aubert:2009ac}
    and Belle~\cite{Ha:2010rf,Sibidanov:2013rkk,Glattauer:2015teq}
    are fitted into a model-independent parametrization
    in terms of $z$-parameter defined as
    $z(q^2,t_{\rm opt})\!=\!(\sqrt{t_+-q^2}-\sqrt{t_+-t_{\rm opt}})/
                          (\sqrt{t_+-q^2}+\sqrt{t_+-t_{\rm opt}})$.
    Here $t_+$ is the threshold $t_+\!=\!(M_B+M_{\pi(D)})^2$
    and a tunable parameter $t_{\rm opt}$ is set to 
    $(M_B+M_{\pi(D)})(\sqrt{M_B}-\sqrt{M_{\pi(D)}})^2$
    to minimize the maximal value of $|z|$.
    We note that lattice data of $f_0$ are also included into the fits
    to make use of the kinematical constraint $f_+(0)\!=\!f_0(0)$.
  }%
  \label{fig:flv:heavy:sld:f+0}
\end{center}
\end{figure}

In the SM,
the $B\!\to\!\pi\ell\nu$ and$ B\!\to\!D\ell\nu$ decays
proceed only through the weak vector current $V_\mu$ due to parity symmetry.
The matrix element for $B\!\to\!\pi\ell\nu$, for instance,
is parametrized by two form factors as 
\be
   \langle \pi(p^\prime) | V_\mu | B(p) \rangle
   = 
   \left\{
      p + p^\prime - \frac{M_B^2-M_\pi^2}{q^2} q
   \right\}_\mu f_+^{B\pi}(q^2)
  +\frac{M_B^2-M_\pi^2}{q^2} q_\mu f_0^{B\pi}(q^2),
   \label{eqn:flv:heavy:ff}
\ee
where $q^2\!=\!(p-p^\prime)^2$ is the momentum transfer
to the lepton pair $\ell\nu$.
For $\ell\!=\!e,\mu$,
the contribution from the ``$f_0$ term'' 
to the differential decays rate $d\Gamma/dq^2$
is suppressed by $m_l^2$.
Hence the experimental value of $|V_{ub}|f_+^{B\pi}(q^2)$
can be obtained from $d\Gamma/dq^2$.
A simultaneous fit to the experimental data
and $f_+^{B\pi}$ from lattice QCD
determines $|V_{ub}|$ as a relative normalization factor.
The same manner is used to determine $|V_{cb}|$ from $B\!\to\!D\ell\nu$
and can apply to alternative determinations discussed below.
We also note that it has been customary to use 
a model-independent parametrization
based on the analyticity of the form factors
to describe their $q^2$ dependence~\cite{Bourrely:1980gp}.
Figure~\ref{fig:flv:heavy:sld:f+0} shows
FLAG's analysis using recent lattice results 
for $B\!\to\!\pi$~\cite{Dalgic:2006dt,Flynn:2015mha,Lattice:2015tia}
and $B\!\to\!D$~\cite{Lattice:2015rga,Na:2015kha}.
While
the state-of-the-art calculations start to achieve good accuracy 
competitive to experiments,
the number of such calculations are still rather limited 
and independent calculations are highly welcome~\cite{Gelzer:2017edb,Colquhoun:2017gfi}.


The analysis of $B\!\to\!D^*\ell\nu$ is more involved,
since it proceeds also through the weak axial current,
and have four form factors.
The previous lattice study~\cite{Bailey:2014tva}
focused the zero recoil limit,
where $d\Gamma/dq^2$ is described by a single form factor in 
\be
   \langle D^*(\epsilon,p^\prime) | A_\mu | B(p) \rangle
   =
   2 i \sqrt{M_B M_{D^*}} \epsilon_\mu \, h_{A_1}
   \hspace{3mm} ({\bf p}\!=\!{\bf p}^\prime\!=\!{\bf 0}).
\ee
Here $\epsilon$ is the polarization vector of $D^*$.
The conventional determination of $|V_{cb}|$
relies on a parametrization based on heavy quark symmetry
to constrain the form factors at non-zero recoils~\cite{Caprini:1997mu}.
It is, however, recently argued
the possibility that uncertainty due to the parametrization
is not fully understood~\cite{Bigi:2017njr,Grinstein:2017nlq,Bernlochner:2017xyx}.
In order to resolve the tension in $|V_{cb}|$,
lattice calculation of all form factors at non-zero recoil is crucial,
and the first preliminary analysis was recently reported
by the Fermilab/MILC Collaboration~\cite{Aviles-Casco:2017nge}.



\begin{figure}[b]%
\begin{center}
  \includegraphics[width=0.45\linewidth,clip]{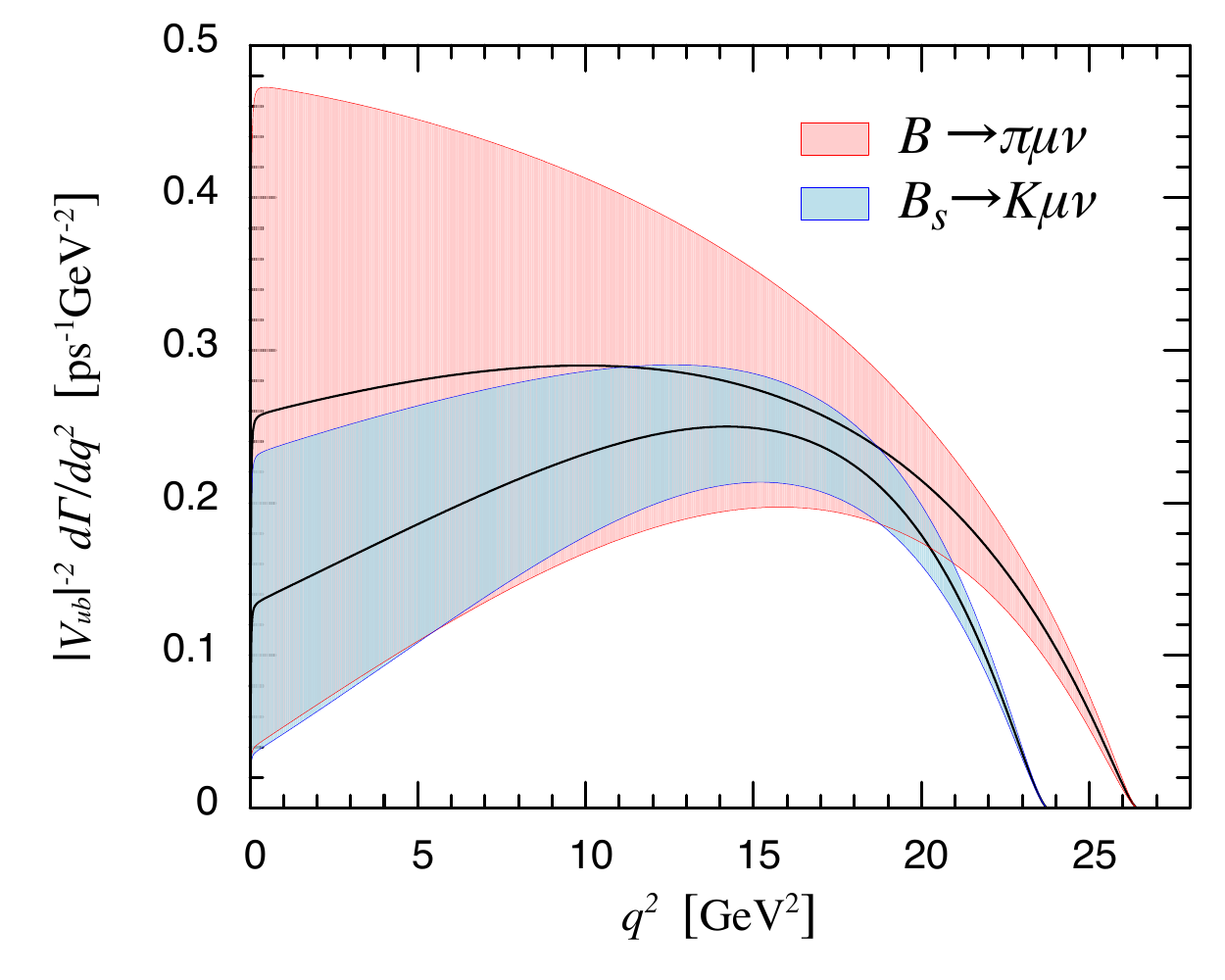}
  \hspace*{1mm}
  \includegraphics[width=0.51\linewidth,clip]{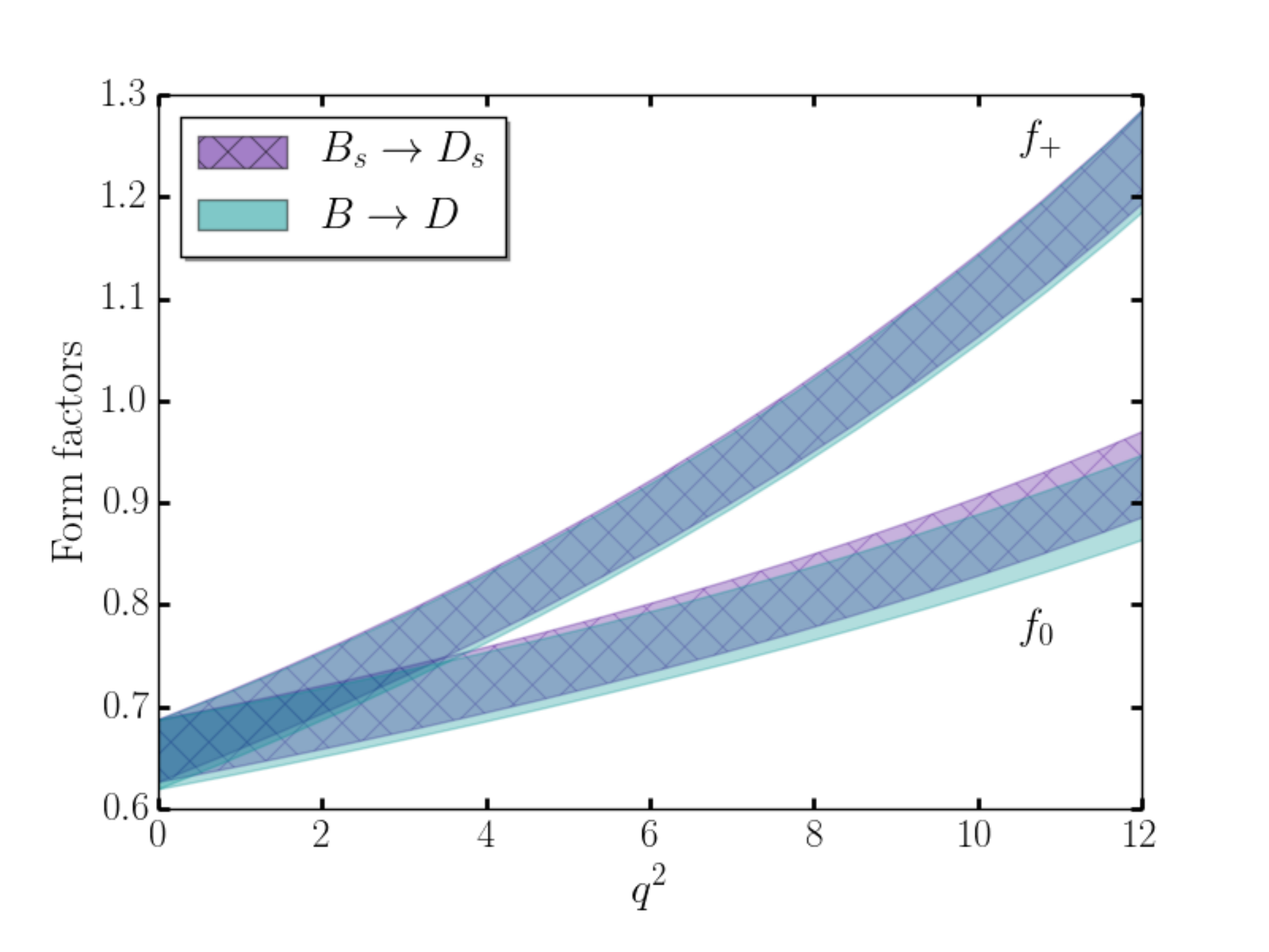}
  \caption{
    Left panel:
    differential decay rates for $B_s\!\to\!K\mu\nu$ (blue band)
    and$ B\!\to\!\pi\mu\nu$ (red band)
    predicted by using recent lattice estimate of relevant form factors
    (figure from Ref.\!\citenum{Flynn:2015mha}).
    Right panel:
    vector ($f_+$) and scalar ($f_0$) form factors
    for $B_s\!\to\!D_s\ell\nu$ (shaded dark blue band)
    and $B\!\to\! D\ell\nu$ (pale blue band) decays
    (figure from Ref.\!\citenum{Monahan:2017uby}).
  }%
  \label{fig:flv:heavy:sld:nonconv}
\end{center}
\end{figure}

Lattice QCD is actively exploring alternative decay modes,
which may elucidate the tension in $|V_{\{ub,cb\}}|$.
Modern calculations are available for
$B_s\!\to\!K\ell\nu$
by the HPQCD~\cite{Bouchard:2014ypa}, RBC/UKQCD~\cite{Flynn:2015mha}
and ALPHA~\cite{Bahr:2016ayy} Collaborations,
and for $B_s\!\to\!D_s\ell\nu$
by the Fermilab/MILC~\cite{Bailey:2012rr}, ETM~\cite{Atoui:2013zza}
and HPQCD~\cite{Monahan:2017uby} Collaborations.
Simulation techniques for $B\!\to\!\{\pi,D^{(*)}\}\ell\nu$ 
can be straightforwardly applied
to attain a similar level of precision
as seen in the right panel of Fig.~\ref{fig:flv:heavy:sld:nonconv}.
The left panel shows that 
the theoretical accuracy is better for $B_s\!\to\!K\ell\nu$ 
than $B\!\to\!\pi\ell\nu$,
since having the (heavier) kaon instead of the pion
reduces the statistical fluctuation and light quark mass dependence
of the form factors.
The latter is advantageous to control the chiral extrapolation.
Therefore, the most crucial issue
for the practical use of these alternative modes
is the experimental feasibility of precise measurements
of $d \Gamma / dq^2$.

\begin{figure}[b]%
\begin{center}
  \parbox{0.41\linewidth}{
    \begin{center}
    \includegraphics[width=0.99\linewidth,clip]{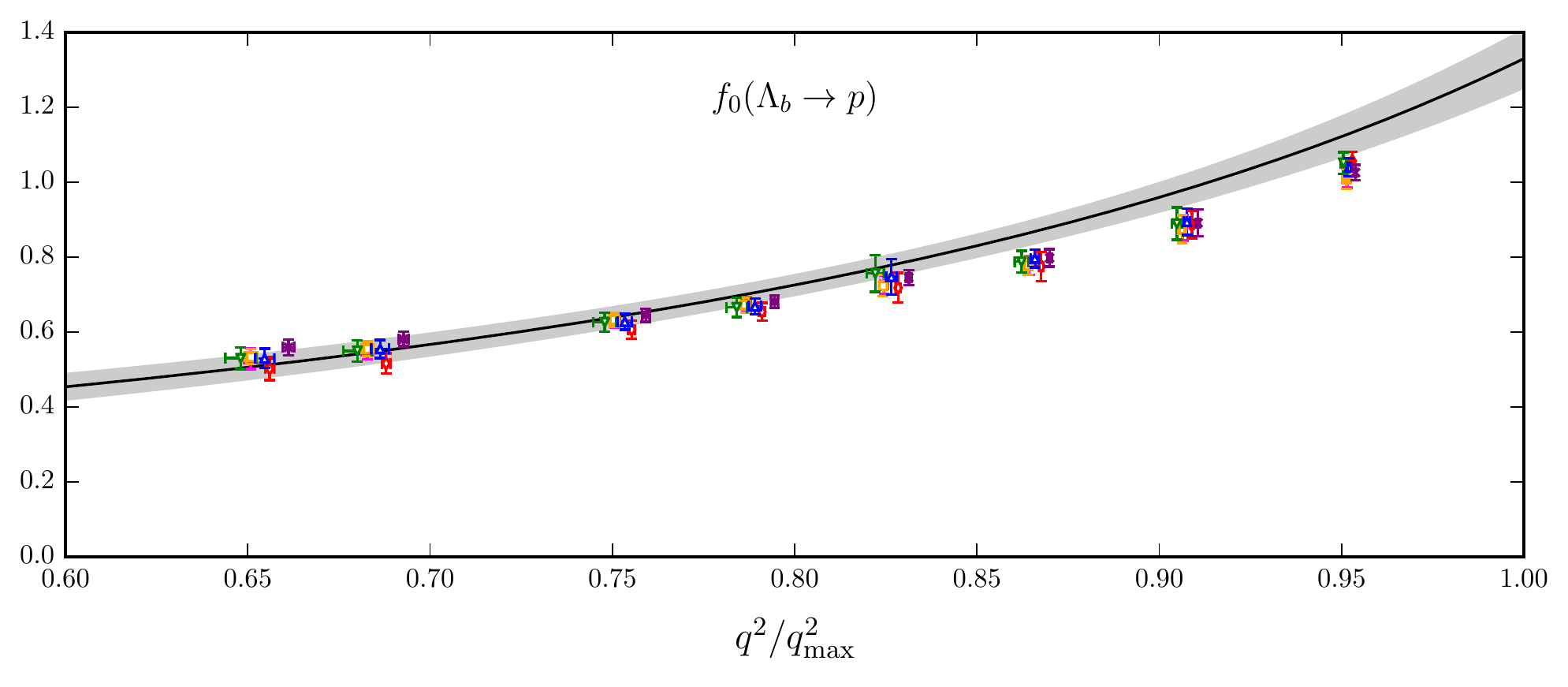}
    \vspace{1mm}
    \includegraphics[width=0.99\linewidth,clip]{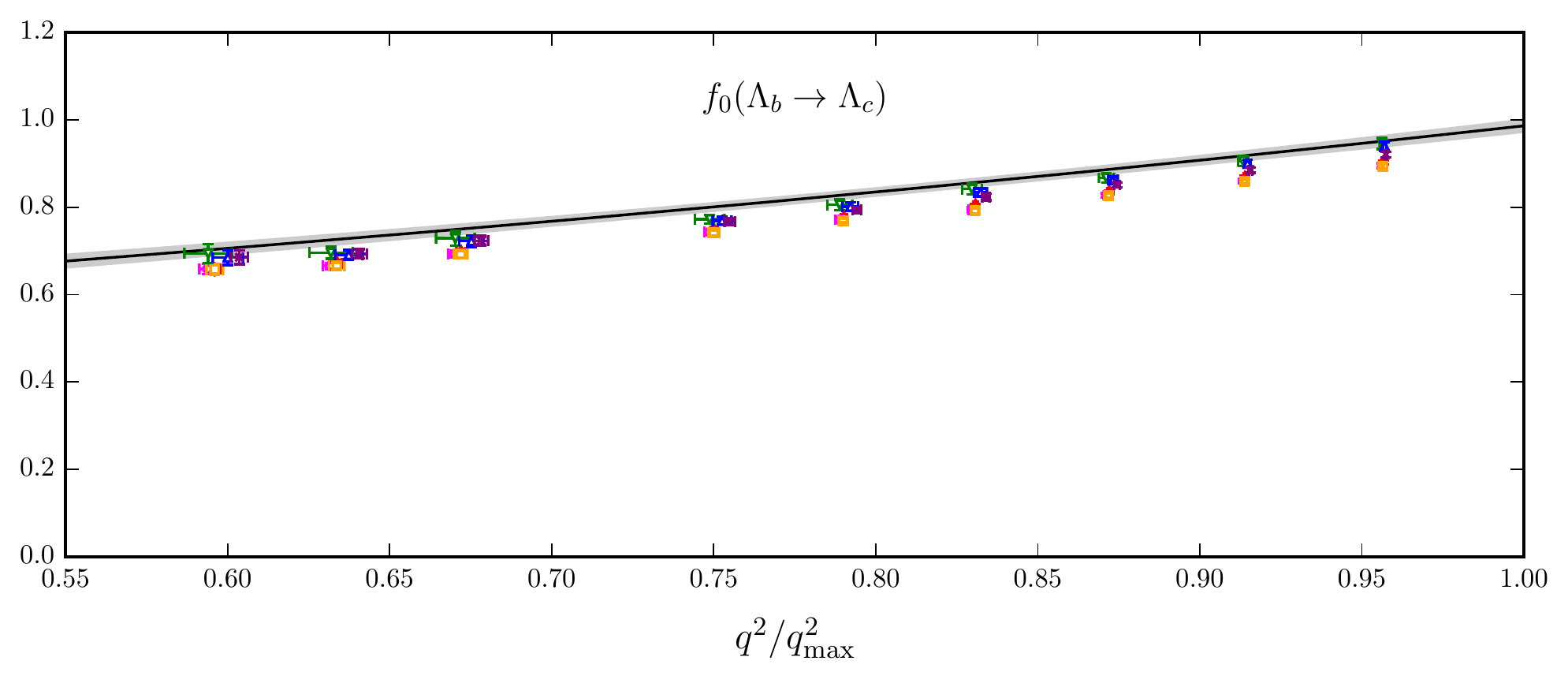}
    \end{center}
  }
  \parbox{0.58\linewidth}{\includegraphics[width=1.0\linewidth,clip]{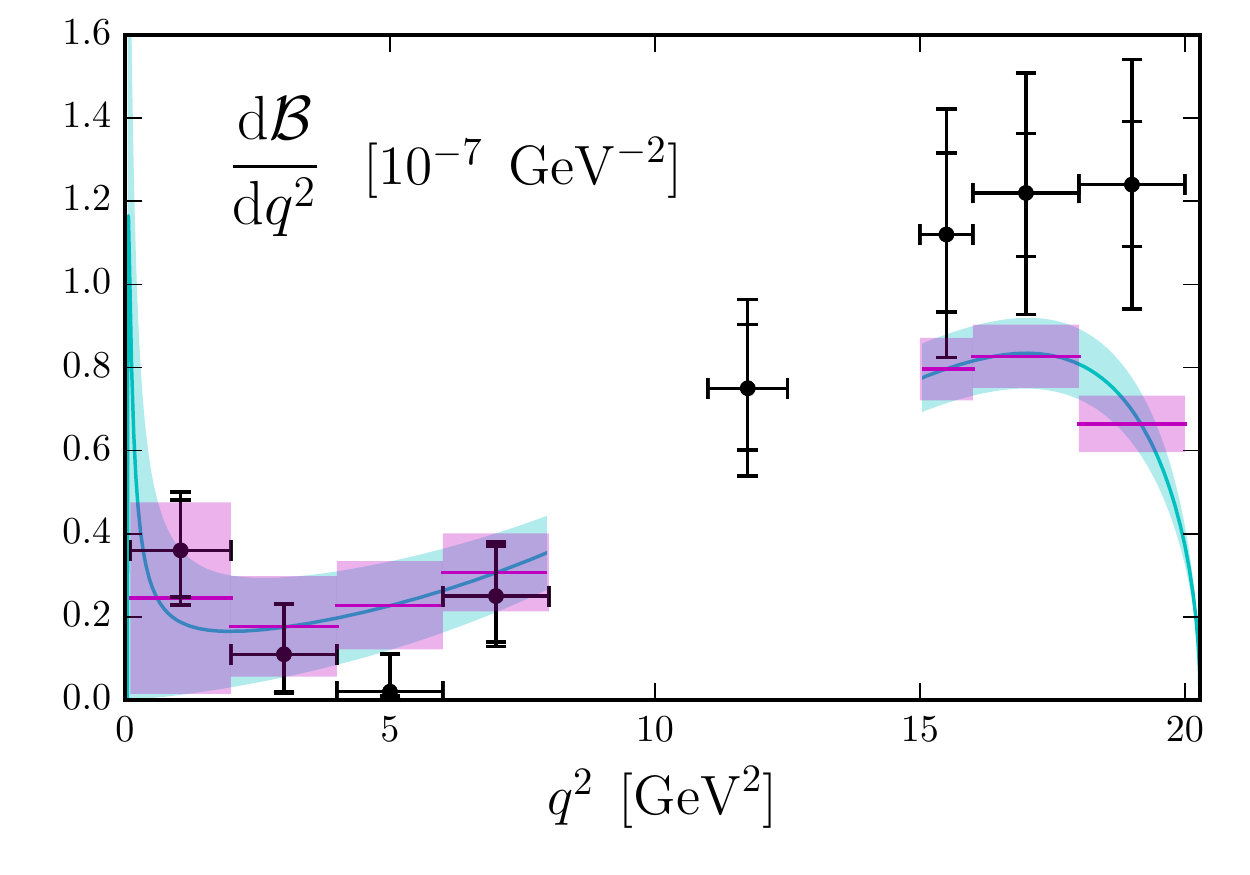}}
  \caption{
    Left panels:
    an example of $\Lambda_b\!\to\!p$ (left top panel)
    and $\Lambda_b\!\to\!\Lambda_c$ (left bottom panel) form factors
    describing matrix elements
    $\langle p\, (\Lambda_c)|V_\mu|\Lambda_b\rangle$
    (figure from Ref.\!\citenum{Detmold:2015aaa}).
    The horizontal axis represents $q^2$
    normalized by $q^2_{\rm max}\!=\!(M_{\Lambda_b}-M_{p (\Lambda_c)})^2$.
    Right panel:
    differential branching fraction of $\Lambda_b\!\to\!\Lambda\ell\ell$
    as a function of $q^2$
    (figure from Ref.\!\citenum{Detmold:2016pkz}).
    The black circle shows the LHCb result~\cite{Aaij:2015xza}
    with two error bars including and excluding the uncertainty
    from the normalization mode $\Lambda_b\!\to\!J/\Psi\Lambda$.
    The continuous blue band is the SM prediction as a function of $q^2$,
    whereas magenta band show the average at each $q^2$ bin.
  }%
  \label{fig:flv:heavy:sld:baryon}
\end{center}
\end{figure}

Baryon decays also provide the alternative determination
of the CKM matrix elements.
The first lattice calculation of
$\Lambda_b\!\to\!p\ell\nu$ and $\Lambda_b\!\to\!\Lambda_c\ell\nu$ form factors
at the physical $b$ quark mass
obtained $|V_{ub}|/|V_{cb}|$ shown in Fig.~\ref{fig:flv:heavy:sld:hint}~\cite{Detmold:2015aaa}.
An example of the relevant form factors is shown in the left panels of
Fig.~\ref{fig:flv:heavy:sld:baryon}.
Recently, the first study of $\Lambda_c\!\to\!\Lambda\ell\nu$ yields
$|V_{cs}|$ consistent with those from $D$ meson decays~\cite{Meinel:2016dqj}.
However baryons are known to be more challenging in controlling systematics,
in particular finite volume effects and chiral extrapolation.
Since the target accuracy for $|V_{\{cb,cs\}}|$ is high,
more thorough study of systematics is recommended
to firmly establish the CKM matrix elements from the baryon decays.


\def\figsubcap#1{\par\noindent\centering\footnotesize(#1)}

 For rare decays, however, 
such theoretical uncertainty could be well below the experimental one.
Reference\!\citenum{Detmold:2016pkz} recently presented
the first calculation of all ten form factors
for $\Lambda_b\!\to\!\Lambda\ell\ell$.
This decay may shed new light on the so-called $B\!\to\!K^*\ell\ell$ anomaly,
namely more than 3\,$\sigma$ tension in its angular distribution~\cite{Descotes-Genon:2013wba},
since these two decays share 
the $b\!\to\!s\ell\ell$ effective weak Hamiltonian.
As shown in the right panel of Fig.~\ref{fig:flv:heavy:sld:baryon},
the SM prediction and the LHCb result~\cite{Aaij:2015xza}
show reasonable consistency for the differential branching fraction.
A simple scenario for the $B\!\to\!K^*\ell\ell$ anomaly, 
namely a negative new physics coupling for an effective interaction
$(\bar{s}_L\gamma_\mu b_L)(\bar{\ell}\gamma^\mu\ell)$, worsens this consistency.
Therefore the authors suggest that the $B\!\to\!K^*\ell\ell$ anomaly is due to
incomplete treatment of the charmonium resonance contribution
($B\!\to\!K^*\psi_n(\to\!ll)$, $\psi_1\!=\!J/\psi$, $\psi_2\!=\!\psi_{2S}$, ...)
in the angular analysis.


\begin{figure}[t]
\begin{center}
  \includegraphics[width=0.4\linewidth,clip]{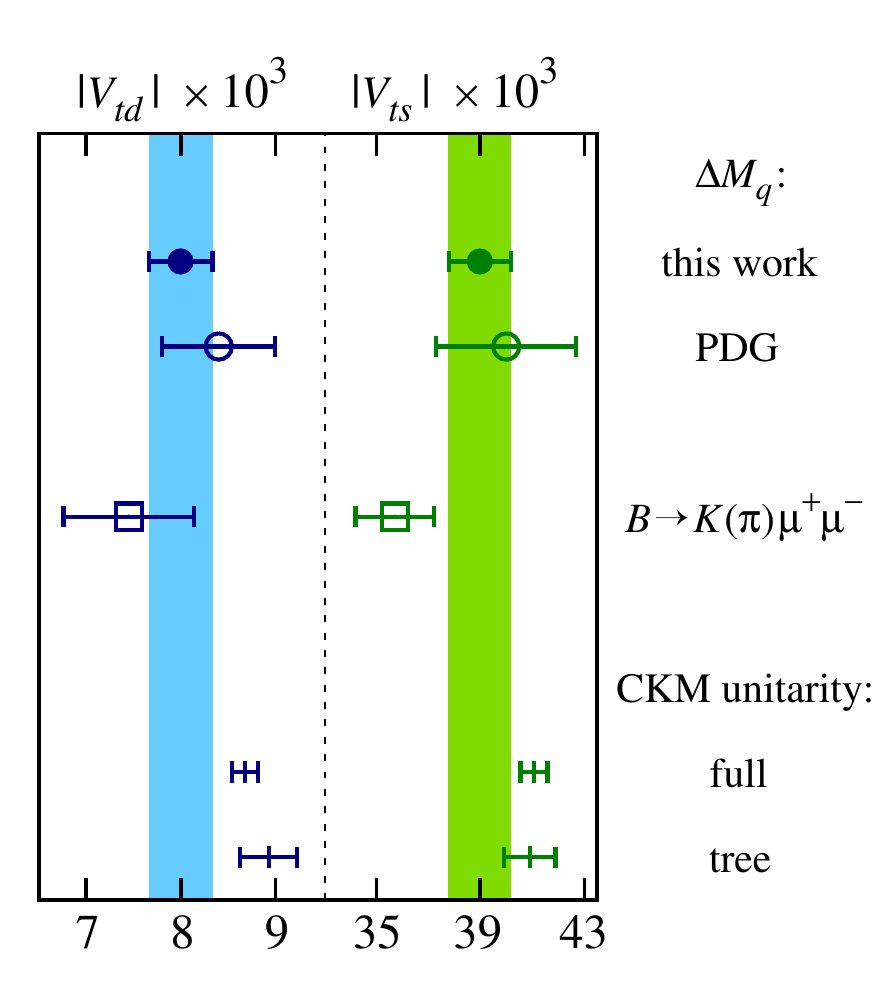}
  \includegraphics[width=0.4\linewidth,clip]{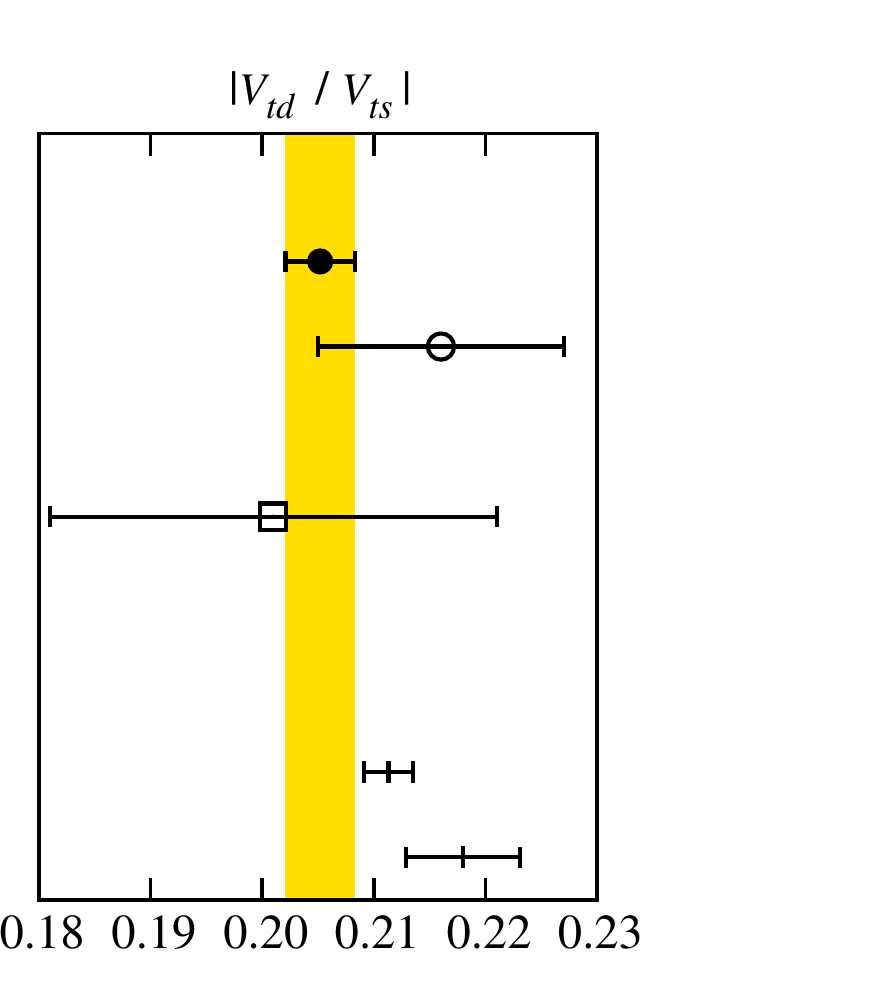}
\end{center}
\caption{
   Comparison of $|V_{td}|$ and $|V_{ts}|$ (left panel)
   and their ratio (right panel)
   (figure from Ref.\!\citenum{Bazavov:2016nty}).
   The filled circles and vertical bands show
   the Fermilab/MILC's estimate
   through the neutral $B_{(s)}$ meson mass difference $\Delta M_q$,
   which shows a 2\,--\,3 times reduction of the uncertainty
   upon the previous estimate (open circles).
   The squares are from the rare decays $B\!\to\!K(\pi)\mu\mu$.
   The plus symbols are obtained by assuming CKM unitarity
   through a global fit to all available inputs
   and a fit limited to inputs from tree-level processes.
}
\label{fig:flv:heavy:mixing}
\end{figure}

For the neutral $B$ meson mixing,
the Fermilab/MILC Collaboration recently published a precise calculation
of all matrix elements in and beyond the SM~\cite{Bazavov:2016nty}.
High statistics, realistic simulation parameters ($a$ and $M_\pi$),
and a better implementation of the renormalization of
four-fermion lattice operators~\cite{Lepage:1992xa,Bazavov:2016nty}
led to 2\,--\,3 times improvement in the relevant CKM elements
as shown in Fig.~\ref{fig:flv:heavy:mixing}.
With such high precision,
a tantalizing tension with CKM unitarity emerges. 
We also note that there is a tension in a matrix element beyond the SM
with ETM Collaborations' result in $N_f\!=\!2$ QCD~\cite{Carrasco:2013zta}.
We hope that independent precision calculations will clarify the source of
these tensions in the near future.


\begin{figure}[h]
\begin{center}
  \includegraphics[width=0.65\linewidth]{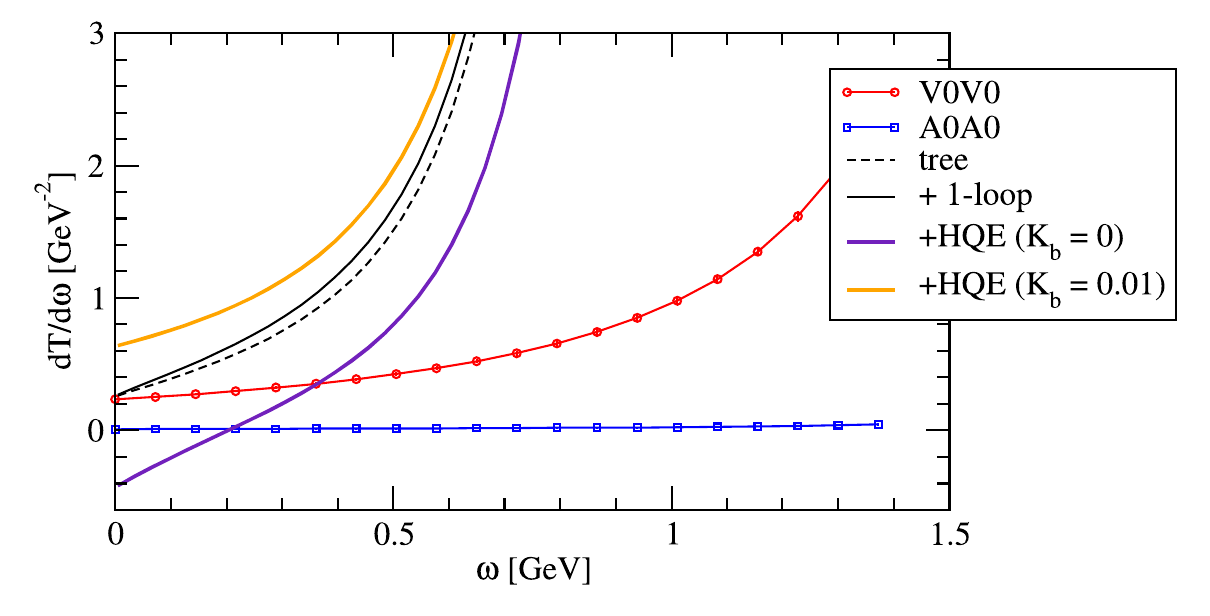}
\end{center}
\vspace{-3mm}
\caption{
   First derivative of the matrix element $T_{\mu\nu}$ as a function of
   kinematical variable $\omega\!=\!M_B-q_0$.
   Red circles and Blue squares 
   are weak vector ($J_\mu\!=\!V_\mu$) and axial vector ($J_\mu\!=\!A_\mu$)
   current contributions, respectively. Here we plot $dT_{\mu\nu}/d\omega$ 
   instead of $T_{\mu\nu}$ itself to avoid contamination from contact terms.
   Thin black dashed line shows the estimate from the heavy quark expansion
   at the leading order in $\alpha_s$ and $1/m_b$,
   whereas the one-loop correction is included into the black solid line.
   The thick lines take account of the $1/m_b$ correction
   with two choices of the non-perturbative input
   $K_b\!=\!-\langle B | \bar{b} (iD)^2 b | B \rangle / (2m_b^2)$~\cite{Manohar:1993qn}.
}
\label{fig:flv:heavy:inclusive}
\end{figure}


While generalization of the Lellouch-Lu\"scher framework is actively pursued,
it is not yet sufficient to allow lattice simulations of 
hadronic $B$ or $D$ decays which have various multi-particle final states.
However, lattice simulation is now being applied to inclusive decays~\cite{Maltman:2017prs,Hashimoto:2017wqo,Liu:2017lpe,Hansen:2017mnd},
which are summed over all possible hadronic final states.
For the hadronic $\tau$ decays $\tau\!\to\!X_s\nu$, for instance,
the optical and Cauchy theorems
relate a normalized decay rate $R_s$ to a contour integral
of the weak current correlator over the complex energy variable $s$
\be
   R_s
   = \frac{\Gamma(\tau\!\to\!X_s \nu_\tau)}
          {\Gamma(\tau\!\to\!e\bar{\nu}_e\nu_\tau)}
   = -\frac{1}{2\pi i} |V_{us}|^2
      \oint_{|s|=s_0} ds\, w(s) \langle 0 | J_\mu J_\mu | 0 \rangle.
\ee
Here, 
$s_0$ and $w(s)$ are appropriately chosen parameter and weight function,
and $J_\mu$ represents the relevant weak current.
By employing a pole type weight function
$w(s) = 1/\Pi_{n=1}^{N}(s+Q_n^2)$ $(Q_n^2>0)$,
the integral can be evaluated by the current correlator
at space like points $s\!=\!-Q_1^2,\ldots,-Q_N^2$,
which can be precisely determined on the lattice.
This is RBC/UKQCD's implementation to determine $|V_{us}|$
from the hadronic $\tau$ decays~\cite{Maltman:2017prs}.
They obtain $|V_{us}|\!=\!0.223$\,--\,0.225~\cite{Boyle:2018dwv,Boyle:2018ilm},
which is in good agreement with those from the kaon decays
shown in the left panel of Fig.~\ref{fig:flv:kaon:semi+leptonic}.

Inclusive semileptonic $B$ decays are more involved but important
application to resolve the discrepancy in $|V_{\{ub,cb\}}|$
from the exclusive decays~\cite{Hashimoto:2017wqo}.
Through the optical theorem,
the hadronic part of the inclusive decay amplitude can be related to
a forward scattering matrix element
\be
   T_{\mu\nu}
   =
   i \int d^4x e^{-iqx} \frac{1}{2M_B}
               \langle B | T\left[ J_\mu^\dagger(x) J_\nu(0) \right] | B \rangle.
\ee     
A preliminary analysis for $B\!\to\!X_c\ell\nu$ at zero recoil
has been reported by Hashimoto {\it et al.}
in Ref.\!\citenum{Hashimoto:2018gld}.
Figure~\ref{fig:flv:heavy:inclusive} compares
the matrix element $T_{\mu\nu}$ between lattice QCD
and the heavy quark expansion (HQE)
used in the conventional inclusive determination of $|V_{cb}|$.
They are reasonably consistent in the perturbative region
of a kinematical variable $\omega\!=\!M_B-q_0\!\sim\!0$.
However, they deviate from each other towards large $\omega$,
where the perturbation series is reasonably convergent,
but the $1/m_b$ correction has substantial uncertainty
due to the choice of the non-perturbative input. 
Further quantitative test of the HQE
and detailed comparison between the inclusive and exclusive
determinations on the same lattice
may give us a hint to resolve the tension in $|V_{\{ub,cb\}}|$.


\def\figsubcap#1{\par\noindent\centering\footnotesize(#1)}
\begin{figure}[b]%
\begin{center}
  \parbox{0.18\linewidth}{\includegraphics[width=1.0\linewidth]{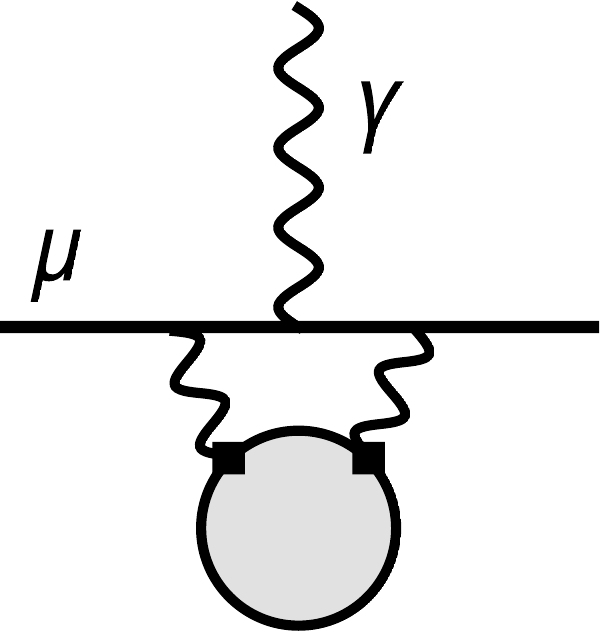}\figsubcap{a}}
  \hspace{1mm}
  \parbox{0.18\linewidth}{\includegraphics[width=1.0\linewidth]{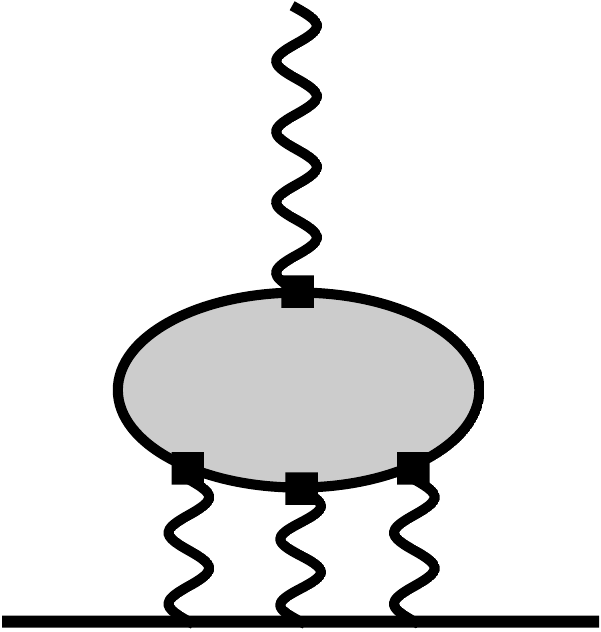}\figsubcap{b}}
  \hspace{1mm}
  \parbox{0.18\linewidth}{\includegraphics[width=1.0\linewidth]{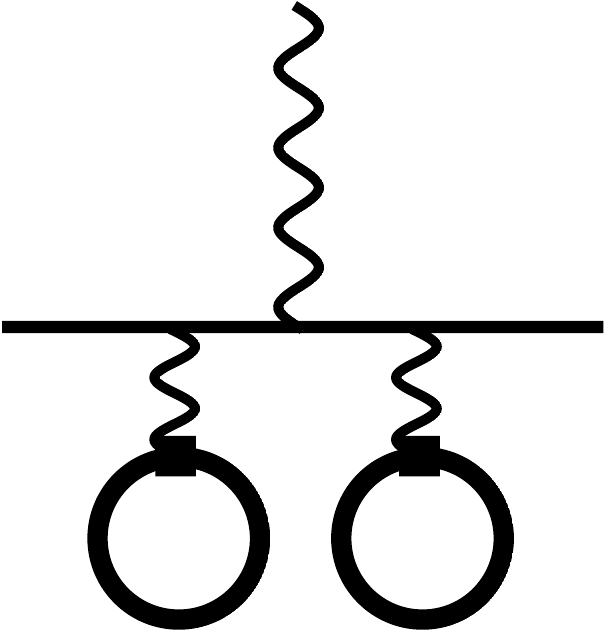}\figsubcap{a-1}}
  \hspace{1mm}
  \parbox{0.18\linewidth}{\includegraphics[width=1.0\linewidth]{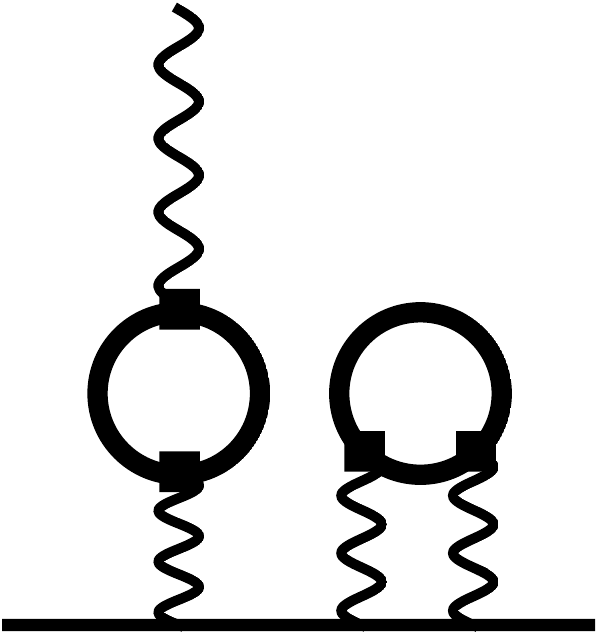}\figsubcap{b-1}}
  \hspace{1mm}
  \parbox{0.18\linewidth}{\includegraphics[width=1.0\linewidth]{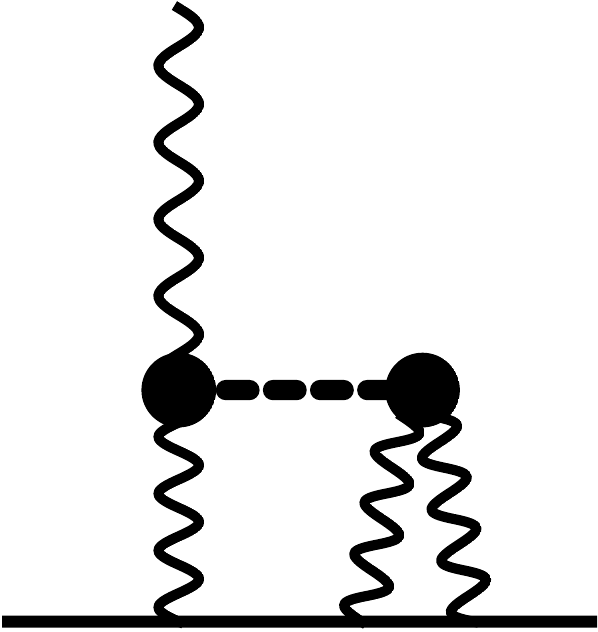}\figsubcap{b-2}}
  \caption{
    The leading HVP (a) and HLbL (b) contributions to $a_\mu$.
    Wavy and solid lines represent the photon and muon propagators,
    respectively.
    The shaded circle for the HVP (HLbL) is the two- (four-)point function
    of the quark EM current $J_\mu^{\rm EM}$ (squares) in QCD.
    (a-1) An example of quark-disconnected diagrams for the HVP.
          Two quark loops (thick lines) are connected by gluons.
    (b-1) An example of leading disconnected diagrams for the HLbL.
          We note that a quark loop with a single $J_\mu^{\rm EM}$ vertex
          vanishes in the SU(3) limit, since the up, down and strange quark
          charges sum up to zero.
    (b-2) The $\pi^0$ contribution to the HLbL. The thick dashed line and
          the solid circles represent the $\pi^0$ propagator
          and $\pi^0\!\to\!\gamma^*\gamma^*$ transition form factor,
          respectively.
  }%
  \label{fig:g-2:contribu}
\end{center}
\end{figure}

\section{Muon anomalous magnetic moment}
\label{sec:g-2}


The past several years have witnessed 
rapid progress in calculating 
the muon anomalous magnetic moment $a_\mu\!=\!(g-2)/2$
on the lattice. 
This quantity is known to great precision $\approx$~0.5\,ppm
both in the SM~\cite{Hagiwara:2011af,Jegerlehner:2017lbd,Davier:2017zfy}
and experiment~\cite{Bennett:2006fi}.
Their 3\,--\,4\,$\sigma$ deviation may be a signal of new physics.
New experiments,
E989 at Fermilab~\cite{Grange:2015fou} and E34 at J-PARC~\cite{Otani:2015jra},
aim to improve the experimental accuracy by a factor of four.
On the theory side,
the accuracy is currently limited by 
hadronic uncertainties coming from 
the hadronic vacuum polarization (HVP) and hadronic light-by-light (HLbL)
contributions shown in Fig.~\ref{fig:g-2:contribu}.
The former has been most precisely calculated from the
dispersive analysis~\cite{Hagiwara:2011af,Jegerlehner:2017lbd,Davier:2017zfy}
of experimental data of 
the so-called $R$-ratio
$\sigma(e^+e^-\!\to\!\mbox{hadrons})/\sigma_{\rm tree}(e^+e^-\!\to\!\mu^+\mu^-)$.
A similar dispersive approach is
under development~\cite{Colangelo:2014dfa,Pauk:2014jza,Colangelo:2017qdm},
and model estimates are currently quoted for the HLbL~\cite{Bijnens:1995xf,Hayakawa:1995ps,Melnikov:2003xd}.
An ultimate goal of recent lattice efforts is
to provide the first-principle prediction for these hadronic contributions
with reduced uncertainty.
 


The leading order HVP contribution $a_\mu^{\rm HVP}$ is expressed
as an integral over the Euclidean momentum $Q^2$~\cite{Lautrup:1969fr,Blum:2002ii}
\be
  a_\mu^{\rm HVP}
  =
  4 \alpha^2
  \int_0^\infty dQ^2 K(Q^2,m_\mu^2) \hat{\Pi}(Q^2),
  \hspace{3mm}
  \hat{\Pi}(Q^2) = \Pi(Q^2) - \Pi(0),
  \label{eqn:g-2:amu:Q2}
\ee
where $K$ is a known weight.
In pioneering works~\cite{Blum:2002ii,Gockeler:2003cw,Aubin:2006xv,Feng:2011zk},
the vacuum polarization function $\Pi(Q^2)$ was
straightforwardly calculated from the EM current correlator
\be  
  \left(Q_\mu Q_\nu - \delta_{\mu\nu}Q^2 \right) \Pi(Q^2)
  =
  \int d^4x e^{iQx} \langle J_\mu^{\rm EM}(x) J_\nu^{\rm EM}(0) \rangle.
  \label{eqn:g-2:amu:vpf}
\ee
%
%
It was later suggested~\cite{Bernecker:2011gh,Feng:2013xsa,Francis:2013qna} that
$\hat{\Pi}(Q^2)$ can be expressed using the correlator 
$G(t)\!=\!\sum_{\bf x}\langle J_\mu^{\rm EM}({\bf x},t) J_\mu^{\rm EM}(0) \rangle$
with zero spatial momentum,
which is less noisy than the right-handed side of (\ref{eqn:g-2:amu:vpf}).
$a_\mu^{\rm HVP}$ is written as an integral over the Euclidean time
\be
  a_\mu^{\rm HVP}
  =
  4 \alpha^2
  \int_0^\infty dt \tilde{K}(t,m_\mu^2) G(t),
  \label{eqn:g-2:amu:tmr}
\ee
which is often employed in recent lattice studies~\cite{DellaMorte:2017dyu,Borsanyi:2017zdw,Blum:2018mom}.
%
%
We note that the integral (\ref{eqn:g-2:amu:Q2}) receives large contribution
from the infrared regime $Q^2\!\approx\!O(m_\mu^2)$.
References~\citenum{Chakraborty:2014mwa,Chakraborty:2015ugp,Chakraborty:2016mwy}
employ another strategy,
which reconstructs $\hat{\Pi}(Q^2)$ in the infrared region
by using its derivatives at $Q^2\!=\!0$ calculated from time moments
$G_{2n}\!=\!\sum_{t}t^{2n}G(t)\!=\!
(-1)^n\left(\partial^{2n} Q^2\hat{\Pi}(Q^2)/\partial Q^{2n} \right)_{Q^2\!=\!0}$.

\begin{figure}[t]%
\begin{center}
  \includegraphics[width=0.478\linewidth,clip]{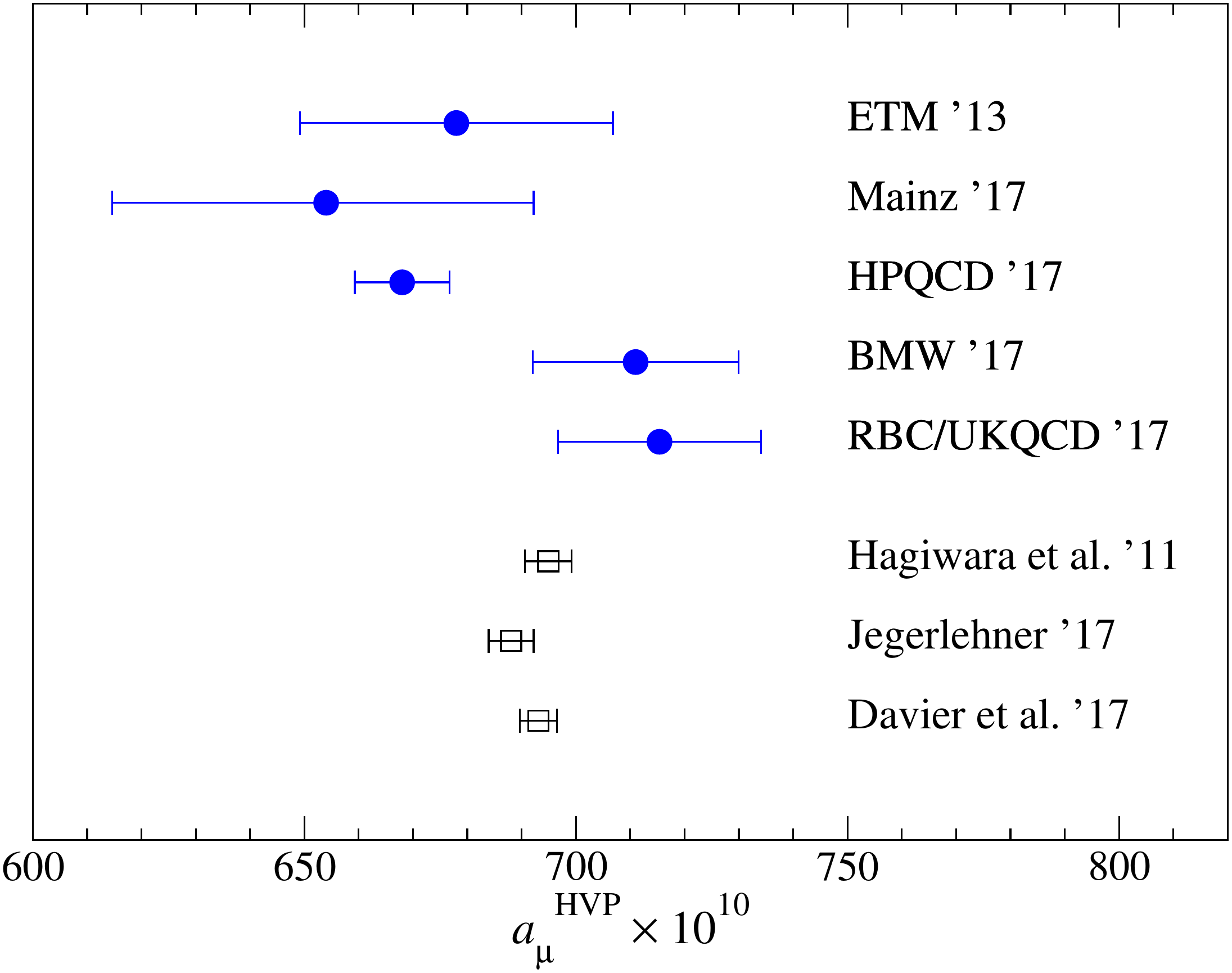}
  \hspace*{1mm}
  \includegraphics[width=0.492\linewidth,clip]{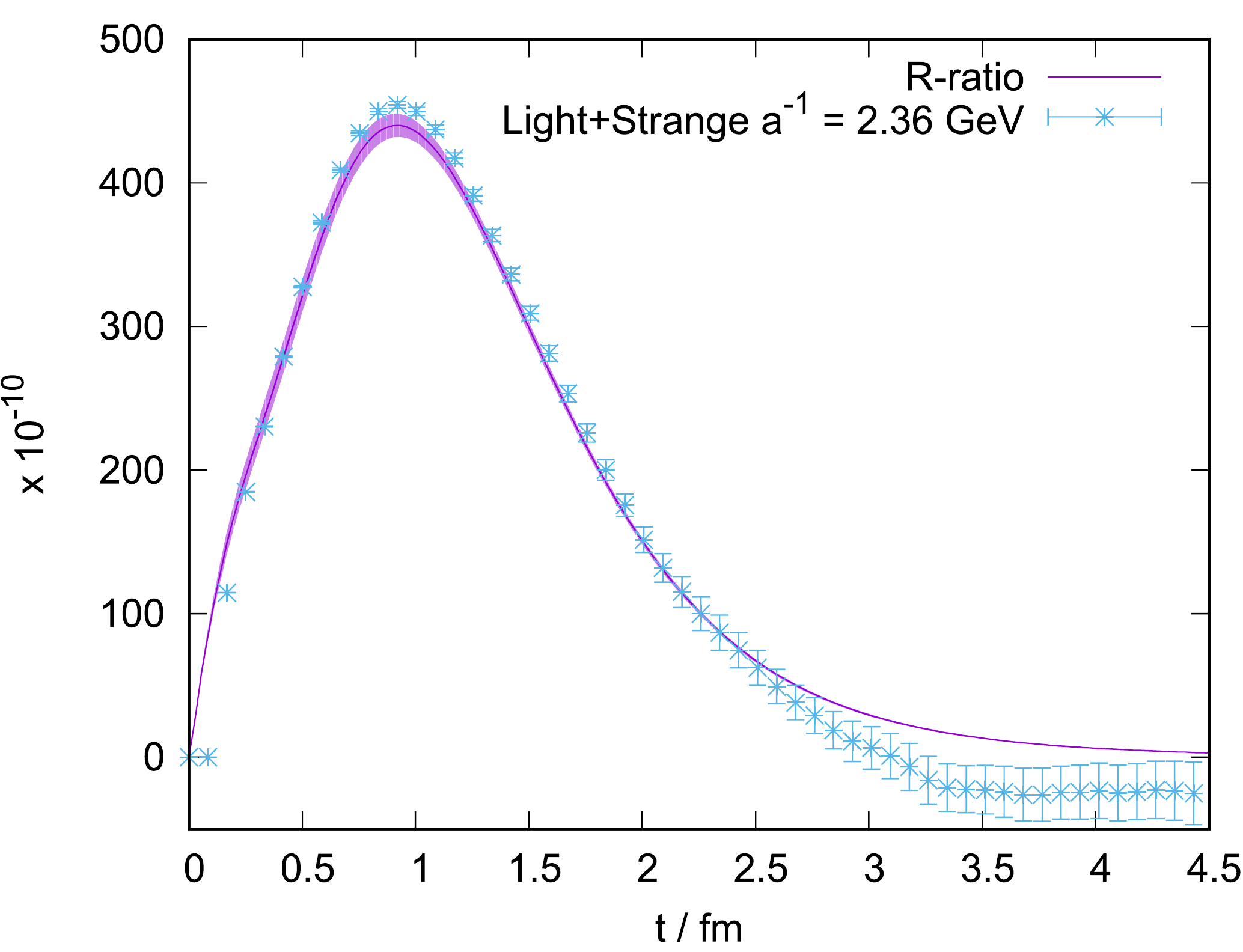}
  \caption{
    Left panel:
    recent lattice estimates of $a_\mu^{\rm HVP}$ (blue solid circles)~\cite{Burger:2013jya,DellaMorte:2017dyu,Chakraborty:2017tqp,Borsanyi:2017zdw,Blum:2018mom}.
    We also plot $a_\mu^{\rm HVP}$ from the dispersive method~\cite{Hagiwara:2011af,Jegerlehner:2017lbd,Davier:2017zfy}.
    Right panel:
    comparison of integrand $\hat{K}(t,m_\mu^2)G(t)$ in 
    Eq.(\protect\ref{eqn:g-2:amu:tmr})
    (figure from Ref.\!\citenum{Lehner:2017kuc}).
    The crosses are lattice data at a finite lattice spacing
    $a\!\sim\!0.08$~fm, 
    whereas the band is obtained from the experimental data of the $R$-ratio.
  }%
  \label{fig:flv:g-2:hvp}
\end{center}
\end{figure}


The challenge in the lattice calculation of $a_\mu^{\rm HVP}$
is to control all uncertainties below 1\,\%
to compete with the current best estimate from the $R$-ratio.
The dominant contribution comes from the connected diagram
with the light quark current
$(2/3)\bar{u}\gamma_\mu u -(1/2)\bar{d}\gamma_\mu d$.
It has been calculated in the isospin limit
with controlled continuum and chiral extrapolations~\cite{Feng:2011zk,Burger:2013jya,Chakraborty:2016mwy,DellaMorte:2017dyu,Borsanyi:2017zdw,Blum:2018mom}, whereas 
finite volume effects are corrected by employing effective field theories
or directly examined by simulating multiple volumes.
Possible corrections have been studied:
strong isospin breaking~\cite{Blum:2018mom,Chakraborty:2017tqp}, 
EM correction~\cite{Giusti:2017jof,Blum:2018mom}, 
disconnected diagrams (Fig.~\ref{fig:g-2:contribu} (a-1))~\cite{Blum:2015you,Chakraborty:2015ugp,DellaMorte:2017dyu,Borsanyi:2017zdw},
strange and charm quark contributions~\cite{Chakraborty:2014mwa,Blum:2016xpd,DellaMorte:2017dyu,Giusti:2017jof,Borsanyi:2017zdw}.
It is confirmed that bottom quarks have small effects~\cite{Colquhoun:2014ica}.
Thanks to these substantial efforts,
the current lattice accuracy for $a_\mu^{\rm HVP}$ has reached
a few\,\% level as summarized in Fig.~\ref{fig:flv:g-2:hvp}~\footnote{
   Here we quote results from Refs.\!\citenum{Chakraborty:2017tqp,Borsanyi:2017zdw,Blum:2018mom} published after this symposium. Their preliminary results
  had been available beforehand at
  {\it the 35th International Symposium on Lattice Field Theory
       (Lattice 2017)}.
}.



Currently,
a largest uncertainty is the statistical fluctuation,
though $G(t)$ is less noisy 
compared to the multi-particle correlators discussed
in Sec.~\ref{sec:spectrum} and
three- and four-point functions in Sec.~\ref{sec:flavor}.
A better control in the infrared region,
namely small $Q^2$ and large $t$, is a crucial issue
towards a more precise determination
and is being actively studied~\cite{Aubin:2012me,Golterman:2014wfa,Borsanyi:2016lpl,DellaMorte:2017dyu,Chakraborty:2016mwy,Izubuchi:2018ecf}.
Another interesting possibly is to combine $G(t)$ on the lattice 
and experimental $R$-ratio data.
The time integral (\ref{eqn:g-2:amu:tmr}) is decomposed into
$a_\mu^{\rm HVP}\!=\!a_{\mu,\rm SD}^{\rm HVP}+a_{\mu,\rm ID}^{\rm HVP}+a_{\mu,\rm LD}^{\rm HVP}$.
The $R$-ratio is then used to evaluate the short ($a_{\mu,\rm SD}^{\rm HVP}$)
and long ($a_{\mu,\rm LD}^{\rm HVP}$) distance contributions
to avoid possibly large discretization effects
and statistical fluctuation, respectively.
In the intermediate region,
$a_{\mu,\rm ID}^{\rm HVP}$ can be calculated from $G(t)$.
This combined analysis in Ref.\!\citenum{Blum:2018mom}
demonstrates good consistency in the integrand of Eq.~(\ref{eqn:g-2:amu:tmr})
between lattice and experimental data (Fig.~\ref{fig:flv:g-2:hvp}),
and led to a most precise estimate
$a_\mu^{\rm HVP}\!=\!692.5(2.7)\!\times\!10^{-10}$.
Simulating finer lattices and better control of the statistical accuracy
can expand the intermediate $t$ window,
and will eventually lead
to a pure theoretical estimate of $a_\mu^{\rm HVP}$ with reduced uncertainty.




There has also been remarkable progress in the lattice calculation of 
$a_\mu^{\rm HLbL}$.
Theoretical calculation thereof is challenging,
because the relevant four-point function of $J_\mu^{\rm EM}$
involves many connected and disconnected diagrams.
There have recently been two approaches
to calculate presumably dominant contributions.
%
Reference~\citenum{Blum:2016lnc} focuses on the connected and
leading disconnected diagrams which survive in the SU(3) limit
(Fig.~\ref{fig:g-2:contribu} (b-1)).
With algorithmic improvements~\cite{Blum:2015gfa},
they obtain statistically significant estimate
$a_\mu^{\rm HLbL}\!=\!5.4(1.4)\!\times\!10^{-10}$
at the physical point $M_{\pi,\rm phys}$ but at a single lattice spacing.
%
Another approach in Ref.~\citenum{Gerardin:2016cqj}
estimates presumably dominant $\pi^0$ contribution (Fig.~\ref{fig:g-2:contribu} (b-2))
through the lattice calculation of the $\pi^0\!\to\!\gamma^*\gamma^*$
form factor,
and yields $a_\mu^{\rm HLbL}\!=\!6.5(0.8)\!\times\!10^{-10}$.
While the quote errors are statistical only,
the reasonable agreement between the different approaches is encouraging
and motivates more realistic simulations
and surveys of systematics, particularly
finite volume effects due to massless photons on the lattice~\cite{Blum:2017cer}.


\section{Conclusions}

In this review,
we have presented highlights of recent progress
on hadron spectrum and flavor physics from lattice QCD.
Masses and transition amplitudes 
can be straightforwardly calculated from lattice correlation functions
for hadrons stable under QCD.
Recent realistic simulations can yield deep insight into the nature
of yet-unestablished states as in the case of $\Xi_{cc}$. 
Decay constants, kaon semileptonic form factors and bag parameters
are now calculated with fully controlled uncertainties,
and lie at the precision frontier of lattice QCD,
where isospin corrections are being studied.
The number of the precision calculations is currently rather limited
for heavy hadron form factors and bag parameters.
However, we can expect more independent calculations in the next few years.

Lattice QCD is now ready to study coupled-channel two-body scatterings.
This leads to recent interesting progress on light resonances,
$\sigma$, $\kappa$, $\rho$ and $K^*$ as well as heavy exotics
such as $X(5568)$ and $Z_c(3900)$. 
These studies are however often limited to unphysically heavy pion masses,
which significantly raise thresholds including pions
and may turn resonances into bound-states.
While such studies deepen our understanding of
the existence form of the hadrons,
simulating the physical pion mass is recommended
in order to make a direct comparison with experiment.

There has been continuous progress on $K\!\to\!\pi\pi$
leading to slight tension with experiment,
which is of great phenomenological interest.
General framework to deal with three-particle states
is necessary for hadronic $B$ and $D$ decays and under active development.
However,
it was proposed that inclusive decays can be straightforwardly studied
without such framework. This may offer useful hints to resolve
long-standing tension in $|V_{ub}|$ and $|V_{cb}|$
between the exclusive and inclusive determinations.


\section*{Acknowledgments}

I would like to thank Sinya Aoki, Shoji Hashimoto and Yoichi Ikeda
for informative communications and assistance
in preparing my talk and this manuscript.


\bibliographystyle{lp17_kaneko}
\bibliography{lp17_kaneko}

\end{document}